\documentclass{IEEEtran}
\usepackage{amsfonts}
\usepackage{mathrsfs}
\usepackage{amssymb,amsmath}

\usepackage{algorithm}
\usepackage{algorithmic}
\usepackage{amsmath,amssymb,epsfig,graphics,subfigure}
\usepackage{theorem}
\usepackage{array,color}
\usepackage[compress]{cite}

\theoremheaderfont{\normalfont\bfseries}

\begin{document}

\title{A Framework for Transceiver Designs for Multi-Hop Communications with Covariance Shaping Constraints}

\author{Chengwen Xing, Feifei Gao, and Yiqing Zhou

\thanks{This work was in part supported by the National Natural Science Foundation of China under Grant No. 61421001, 111 Project of China under Grant. B14010, and the Project of China Mobile Research Institute under Grant No. [2014]451. Copyright (c) 2015 IEEE. Personal use of this material is permitted. However, permission to use this material for any other purposes must be obtained from the IEEE by sending a request to pubs-permissions@ieee.org.}

\thanks{C. Xing is with the School of Information and Electronics, Beijing Institute of Technology, Beijing 100081, China (E-mail: chengwenxing@ieee.org). }

\thanks{
F. Gao is with Tsinghua National Laboratory for Information Science and Technology, Tsinghua University, Beijing, China (E-mail: feifeigao@ieee.org).}

\thanks{Y. Zhou is with Institute of Computing Technology, Chinese Academy of Sciences, and Beijing Key Laboratory of
Mobile Computing and Pervasive Devices, Beijing 100190, China (E-mail: zhouyiqing@ict.ac.cn).}
}

\maketitle

\begin{abstract}
For multiple-input multiple-output (MIMO) transceiver designs, sum power constraint is an elegant and ideal model. When various practical limitations are taken into account e.g., peak power constraints, per-antenna power constraints, etc., covariance shaping constraints will act as an effective and reasonable model.
In this paper, we develop a framework for transceiver designs for multi-hop communications under covariance shaping constraints. Particularly, we focus on  multi-hop amplify-and-forward (AF) MIMO relaying communications which are recognized as a key enabling technology for device-to-device (D2D) communications for next generation wireless systems such as 5G. The proposed framework includes a broad range of various linear and nonlinear transceiver designs as its special cases. It reveals an interesting fact that the relaying operation in each hop can be understood as a matrix version weighting operation. Furthermore, the nonlinear operations of Tomolision-Harashima Precoding (THP) and Decision Feedback Equalizer (DFE) also belong to the category of this kind of matrix version weighting operation. Furthermore, for both the cases with only pure shaping constraints or joint power constraints, the closed-form optimal solutions have been derived. At the end of this paper, the performance of the various designs is assessed by simulations.
\end{abstract}


\section{Introduction}
\label{sect:intro}
Multi-hop relaying communications have attracted a lot of attention recently because of both its theoretical and practical importance \cite{D2D1}. From theoretical viewpoint, multi-hop relaying networks include some well-known systems such as dual-hop relaying and point-to-point communication systems as its special cases.  Meanwhile, multi-hop relaying technique is a fundamental technique to enable device-to-device (D2D) communications \cite{D2D1,D2D2}. As it can effectively offload the traffic loads from overloaded macro base stations (BSs) to lightly loaded pico BSs or femto BSs or even small cell BSs, D2D communication technology is envisioned as a key enabling technology to realize high spectrum efficiency for next generation communication systems e.g., 5G wireless systems.

The relaying strategies at relays can be classified into various categories e.g., amplify-and-forward (AF), decode-and-forward (DF), compressed-and-forward (CF) and so on \cite{Jin2010}. In general, each relaying strategy has its own advantages, and which one is the best is really a meaningless question without specific system settings. Due to implementation simplicity and security issue, AF strategies have gained lots of attention. With channel state information (CSI), transceiver designs can greatly improve system performance. Transceiver designs for AF MIMO relaying systems have been extensively studied in the literatures \cite{Yongming2010,WXu2012,Tang07,Schizas07,Medina07,Guan08,Mo09,Rong09,Tseng09,Li09,WeiXu2011}.

When there are multiple data streams are transmitted simultaneously, it is hard to give a dominated performance metric which can be argued better than any another one. Generally speaking, there are various design criteria for transceiver designs for AF MIMO relaying networks. The most widely used criteria are capacity maximization \cite{Tang07,Rong09,Medina07} and data mean-square-error (MSE) minimization \cite{Guan08,Mo09,Tseng09,Rong09}. Capacity reflects how much information can be reliably transmitted, while MSE demonstrates how accurately the desired signals can be recovered at the destination. From the implementation point of view, transceivers designs can be classified into two main categories, i.e., linear transceiver designs and nonlinear transceiver designs. Linear transceiver designs can stricke a balance between performance and complexity \cite{Rong09}. On the other hand, nonlinear transceiver designs can improve bit error rates (BERs) at the cost of high implementation complexity \cite{Sanguinetti2012}.
In the existing works, the nonlinear transceiver designs are usually referred to as the transceivers with Tomolision-Harashima Precoding (THP) at source \cite{THP_Tseng} or Decision Feedback Equalizer (DFE) at destination \cite{Sanguinetti2012}.
Furthermore, taking channel estimation errors into account robust transceiver designs have also attracted lots of attention \cite{Xing10,Xing1012,Yunlong2014,JiaHeng2014,Chalise10,Rong20115,THP_Tseng}. Following this logic, the transceiver designs for multi-hop AF MIMO relaying are investigated in \cite{Rong2009TWC,XingTSP2013,JSAC_Xing2012}, in which both linear and nonlinear transceiver designs are investigated with various performance metrics and even imperfect CSI.

Most of the existing works mainly focus on the transceiver designs with simple and ideal sum power constraints. Unfortunately, there are many practical physical constraints in the practical transceiver designs.
For example, as each antenna has its own power amplifier actually the dynamic range of each amplifier must not exceed a threshold and per-antenna power constraints may be more practical \cite{Palomar2004}. Moreover, sum power constraint is only a definition in the statistical average sense and thus there may be some outages for the specific power constraints at amplifiers. To relieve the outage effects, peak power constraint on the transmitted signal covariance matrix  will be an effective model \cite{Palomar2004,Scaglione2002,Dai2012}.
It is worth noting that $l_{p}$-norm power constraint can also be successfully approximated by joint power constraints consisting of shaping (maximum eigenvalue) and sum power constraints \cite{Feiten2007}. In order to take these constraints in account and still keep graceful closed-form solutions, covariance shaping constraints are usually exploited in the transceiver designs \cite{Palomar2004,Scaglione2002,Dai2012,Feiten2007}. This kind of constraints can effectively mode practical constraints and avoid high complexity numerical computations in the transceiver deigns. In a nutshell, covariance shaping constraints are a kind of useful constraints limiting the transmit power in virtual spatial directions including spectral masks, peak power constraints, per-antenna power constraints and so on \cite{Palomar2004,Scaglione2002,Dai2012}.

In this paper, we take a further step to investigate the transceiver designs for multi-hop cooperative networks under covariance shaping constraints. Both linear transceiver designs and nonlinear transceiver designs are taken into account. In particular, we investigate in depth the transceiver designs with pure shaping constraints and with joint power constraints comprising of sum power constraints and maximum eigenvalue constraints. The main contributions of our work are listed as follows.

\begin{enumerate}

\item The proposed framework includes a wide range of transceiver designs as its special cases e.g., linear transceiver designs with additively Schur-convex/concave objective functions and nonlinear transceiver designs with multiplicatively Schur-convex/concave objective functions. The framework reveals a fact that for the various considered objective functions, in the nature they can always be unified into a multiple objective optimization problem. It is also shown by our framework that the transceiver designs can be decomposed into a series of subproblems which only relate with their respective local CSI.

\item Based on the proposed framework, an interesting and useful understanding of transceiver designs for multi-hop AF MIMO relaying is given. However for AF relaying strategy the noise at each relay will be amplified and forwarded to the next hop, this procedure can be understood as a matrix version weighting operation. Specifically, the AF relaying operation in any hop will act as a matrix weighting operation on the MSE of the remaining successive hops. It may be the reason why AF MIMO relaying looks complicated, but it usually enjoys elegant and simple optimal solutions just as point-to-point MIMO systems. In addition, for nonlinear transceiver designs, the nonlinear operations THP and DFE can also be understood as the same kind of matrix version weighting operation with different matrix version slope and intercept.

\item For the transceiver designs under pure shaping constraints or joint power constraints, the explicit optimal structures of the optimal transceivers can be derived. On the one hand, the transceiver designs under pure shaping constraints have the explicit closed-form optimal solutions which are independent of the specific formulations of objective functions. On the other hand, for the transceiver designs under joint power constraints, based on the optimal structures the remaining variables are only a series of scalar variables that can be efficiently solved by a variant of water-filling solutions named cave water-filling solutions. These structures greatly simplify the practical designs and enable distributed implementation of the proposed algorithm.

\end{enumerate}

The rest of this paper is organized as follows. In Section~\ref{set: system model}, the system model is given and the unified transceiver design under covariance shaping constraints is formulated in Section~\ref{Problem_Formulation}. After that, the considered optimization is simplified into a multi-objective optimization in Section~\ref{Simplified_Optimization}. The optimal solutions for the transceiver designs with pure shaping constraints and joint power constraints are derived in Sections~\ref{Pure_Shaping_Constraint} and~\ref{Joint_Power_Constraint}, respectively. The performance of the different designs is evaluated in Section~\ref{simulation}. Finally, the conclusions are drawn in Section~\ref{Conclusions}.

\noindent \textbf{Notation:} Throughout the whole paper, the following mathematical notations are used. Boldface
lowercase letters denote vectors, and boldface uppercase letters
denote matrices. The notation ${\bf{Z}}^{\rm{H}}$ denotes the
Hermitian of the matrix ${\bf{Z}}$. The symbol ${\rm{Tr}}({\bf{Z}})$ represents the
trace of the matrix ${\bf{Z}}$.  The notation ${\bf{Z}}^{1/2}$ is the
Hermitian square root of the positive semi-definite matrix
${\bf{Z}}$, and it is also a Hermitian matrix. The symbol $\lambda_i({\bf{Z}})$ represents the $i^{\rm{th}}$ largest eigenvalue of ${\bf{Z}}$. For two Hermitian matrices, the equation ${\bf{C}} \succeq
{\bf{D}}$ means that ${\bf{C}}-{\bf{D}}$ is a positive semi-definite
matrix. The symbol ${\boldsymbol \Lambda} \searrow $ represents a rectangular diagonal
matrix with nonincreasing diagonal elements.

\section{System Model}
\label{set: system model}

\begin{figure}[!ht]
\centering
\includegraphics[width=0.5\textwidth]{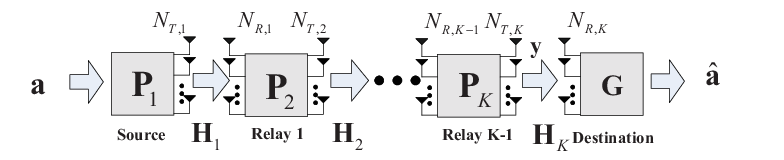}
\caption{A multi-hop MIMO relaying system with linear transceivers.}\label{fig:1}
\end{figure}

In this paper, we are concerned with a multi-hop AF MIMO relaying network. As shown in Fig.~\ref{fig:1}, one source node with $N_{T,1}$ transmit antennas wants to communicate with a destination node with $N_{T,K}$ receive antennas through $K-1$ relay nodes. For the $k^{\rm{th}}$ relay (the $(k+1)^{\rm{th}}$ node), it has $N_{R,k}$ receive
antennas and $N_{T,k+1}$ transmit antennas. In order to guarantee the transmitted data ${\bf{s}}$ can be
recovered at the destination node, it is assumed that
$N_{T,k}$ and $N_{R,k}$ are greater than or equal to $N$ \cite{Guan08}.
 It is straightforward that the a dual-hop/two-hop AF MIMO relaying network is the special case with $K=2$.

At the source node, an $N\times 1$ data vector ${\bf{a}}$ with covariance matrix ${\bf{R}}_{\bf{a}} = {\mathbb{E}}\{{\bf{a}}{\bf{a}}^{\rm H}\} = \sigma_{\bf{a}}^2{\bf{I}}_N$ is transmitted. It should be highlighted that in our work, both linear and nonlinear transmitters are taken into account. For nonlinear transmitters, before going through the precoder matrix ${\bf{P}}_1$ at the source, the vector ${\bf{a}}$ may be preprocessed first. As a result, the signal finally transmitted by the source is denoted by ${\bf{x}}_0$ instead of ${\bf{a}}$ and its specific formulas will be discussed later. The received signal ${\bf{x}}_1$ at the first relay is ${\bf{x}}_1= {\bf{H}}_{1}{\bf{P}}_1{\bf{x}}_0+{\bf{n}}_1$
where ${\bf{H}}_{1}$ is the MIMO channel matrix between the source and the first relay. In addition, ${\bf{n}}_1$ is the additive Gaussian noise vector at the first relay with mean zero and
covariance matrix ${\bf{R}}_{n_1}=\sigma_{n_1}^2{\bf{I}}_{N_{R,1}}$.

The received signal ${\bf{x}}_1$ at the first relay node is first multiplied by a forwarding matrix ${\bf{P}}_2$ and then the resultant signal is
transmitted to the second relay node. The received signal ${\bf{x}}_2$ at
the second relay node is ${\bf{x}}_2={\bf{H}}_{2}{\bf{P}}_2{\bf{x}}_1+{\bf{n}}_2$, where ${\bf{H}}_{2}$ is the MIMO channel matrix between the first
and the second relay nodes. Additionally, ${\bf{n}}_2$ is the additive Gaussian noise
vector at the second relay with mean zero and covariance matrix ${\bf{R}}_{n_2}=\sigma_{n_2}^2{\bf{I}}_{N_{R,2}}$. Similarly, the received signal at the $k^{\rm{th}}$ relay node can be written as
\begin{align}
{\bf{x}}_{k}={\bf{H}}_{k}{\bf{P}}_{k}{\bf{x}}_{k-1}+{\bf{n}}_{k}
\end{align}where ${\bf{H}}_k$ is the channel for the $k^{\rm{th}}$ hop, and ${\bf{n}}_{k}$ is the additive Gaussian noise with mean zero and covariance matrix ${\bf{R}}_{n_{k}}=\sigma_{n_k}^2{\bf{I}}_{N_{T,k}}$. The received signal covariance matrix ${\bf{R}}_{{\bf{x}}_k}$ at the $k^{\rm{th}}$ relay node satisfies the following recursive formula
\begin{align}
\label{R_x}
{\bf{R}}_{{\bf{x}}_k}&= {\bf{H}}_{k}{\bf{P}}_k{\bf{R}}_{{\bf{x}}_{k-1}}{\bf{P}}_k^{\rm{H}}{\bf{H}}_{k}^{\rm{H}}+{\bf{R}}_{n_k}.
\end{align}

The covariance matrix of the transmitted signal at the $k^{\rm{th}}$ node (including both source and relays) is ${\bf{P}}_{k}{\bf{R}}_{{\bf{x}}_{k-1}}{\bf{P}}_{k}^{\rm{H}}$ and in practice there are naturally several constraints on these covariance matrices. The most widely used constraint is the sum power constraint i.e., ${\rm{Tr}}({\bf{P}}_{k}{\bf{R}}_{{\bf{x}}_{k-1}}{\bf{P}}_{k}^{\rm{H}})\le P_k$. In order to limit the transmit power in virtual spatial directions the covariance shaping constraint on the transmitted signal covariance matrix is formulated as
\begin{align}
{\bf{P}}_{k}{\bf{R}}_{{\bf{x}}_{k-1}}{\bf{P}}_{k}^{\rm{H}} \preceq {\bf{R}}_{{\bf{s}}_k}
\end{align} which includes the following constraints as its special cases \cite{Palomar2004}.

\noindent $\bullet$  Peak power constraints:

Note that sum power constraints are defined in the sense of statistical average, but for each power amplifier the power budget limits are deterministic and independent with each other. To shrink the gap between practical phenomena and theoretical model, an effect way is to add peak power constraints \cite{Palomar2004,Scaglione2002,Dai2012}, and therefore we have
\begin{align}
{\bf{P}}_{k}{\bf{R}}_{{\bf{x}}_{k-1}}{\bf{P}}_{k}^{\rm{H}} \preceq \tau_{k,{\max}}{\bf{I}}.
\end{align}

\noindent $\bullet$ Independent Power Constraints Per Antenna:

A simple way to limit each diagonal element of the transmit covariance matrix i.e., $[{\bf{P}}_{k}{\bf{R}}_{{\bf{x}}_{k-1}}{\bf{P}}_{k}^{\rm{H}}]_{i,i}\le P_{i,k}$ is to exploit the following constraint \cite{Palomar2004}
\begin{align}
{\bf{P}}_{k}{\bf{R}}_{{\bf{x}}_{k-1}}{\bf{P}}_{k}^{\rm{H}} \preceq {\rm{diag}}\{\{p_{i,k}\}_{i=1}\}.
\end{align}

\noindent $\bullet$  Spectral mask:

For wire line systems e.g., digital subscriber line (DSL) spectral masks are exploited to guarantee spectral compatibility with different users that share the same cable simultaneously \cite{Palomar2004}.

\noindent $\bullet$  Power constraint along a spatial direction:

Defining the direction by using unitary vector ${\bf{u}}$ the power in this direction equals ${\bf{u}}^{\rm{H}}{\bf{R}}_{{\bf{s}}_k}{\bf{u}}$ and in some cases the leakage power in this direction  should be below a threshold. This constraint can be properly added to transceiver designs by judiciously designing ${\bf{R}}_{{\bf{s}}_k}$. This result is very useful for multiuser communications and mutual interference coordination.

\subsection{Linear Transceiver}
When the linear transceivers are deployed by the relaying networks as shown in Fig.~\ref{fig:1}, at the source the transmitted signal satisfies ${\bf{x}}_0={\bf{a}}$ and the received signal at the destination is
\begin{align}
\label{received_signal_linear}
{\bf{r}} = [{\prod_{k=1}^K}{\bf{H}}_{k}{\bf{P}}_k]{\bf{a}}  + \sum_{k=1}^{K-1}\{ [\prod_{l={k+1}}^K{\bf{H}}_{l}{\bf{P}}_l]{\bf{n}}_k
\}+{\bf{n}}_K,
\end{align}where ${\prod_{k=1}^K}{\bf{Z}}_k$ denotes ${\bf{Z}}_K\times \cdots \times {\bf{Z}}_1$. Meanwhile, at the destination a linear equalizer ${\bf{G}}$ is adopted to recover the desired signal and the data detection mean square error (MSE) matrix is derived to be
\begin{align}
\label{MSE_Matrix_1}
{\boldsymbol {\Phi}}_{\rm{MSE}}({\bf{G}},\{{\bf{P}}_k\}_{k=1}^K)=
\mathbb{E}\{({\bf{G}}{\bf{r}}-{\bf{a}})({\bf{G}}{\bf{r}}-{\bf{a}})^{\rm{H}}\},
\end{align}where the expectation is taken with respect to random data and noises.

\subsection{Decision Feedback Equalizer}

\begin{figure*}[!ht]
\centering
\includegraphics[width=0.64\textwidth]{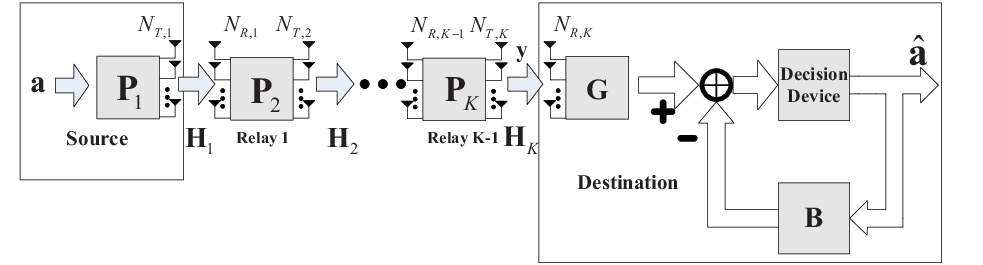}
\caption{A multi-hop MIMO relaying system with DFE at destination.}\label{fig:2}
\end{figure*}

When decision feedback equalizer (DFE) is adopted at the destination and linear precoding is used at the source as shown in Fig.~\ref{fig:2}, the transmitted signal at the source is still ${\bf{a}}$ and the received signal at the destination is the same as (\ref{received_signal_linear}). While the
desired signals are recovered through a DFE and the output signal equals \cite[P.447]{Palomar2007}
\begin{align}
{\bf{y}} = \{{\bf{G}}[{\prod_{k=1}^K}{\bf{H}}_{k}{\bf{P}}_k]-{\bf{B}}\}{\bf{a}}  + {\bf{G}}\{ \sum_{k=1}^{K-1}\{ [\prod_{l={k+1}}^K{\bf{H}}_{l}{\bf{P}}_l]{\bf{n}}_k
\}+{\bf{n}}_K\},
\end{align}where ${\bf{B}}$ is a strictly lower triangular matrix.  With DEF the data detection MSE matrix equals
\begin{align}
\label{MSE_Matrix_2}
{\boldsymbol {\Phi}}_{\rm{MSE}}({\bf{G}},\{{\bf{P}}_k\}_{k=1}^K,{\bf{B}})=
\mathbb{E}\{({\bf{y}}-{\bf{a}})({\bf{y}}-{\bf{a}})^{\rm{H}}\}.
\end{align}

\begin{figure*}[!ht]
\centering
\includegraphics[width=0.64\textwidth]{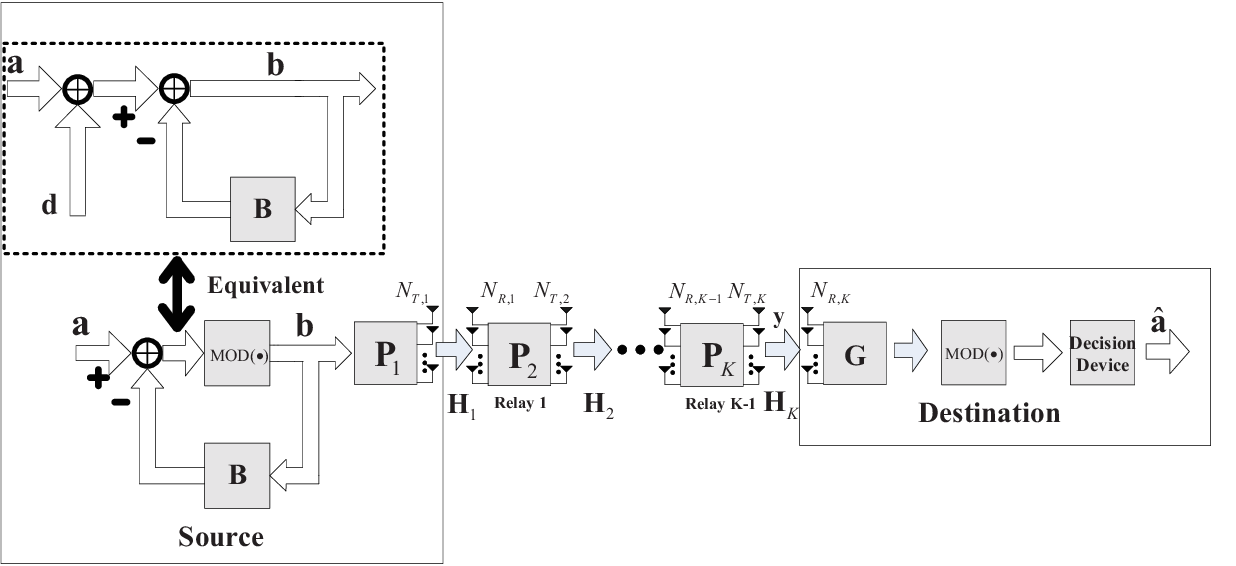}
\caption{A multi-hop MIMO relaying system with THP at source.}\label{fig:3}
\end{figure*}

\subsection{Tomlinson-Harashima
Precoding}
On the other hand, according to dirty paper coding (DPC) mutual interference can be precanceled by exploiting Tomlinson-Harashima
Precoding (THP)  at the source. As shown in Fig.~\ref{fig:3},  at the transmitter, before sending out the data vector ${\bf{a}}$ is fed into the a
precoding unit comprising of a $N\times N$ feedback matrix ${\bf{B}}$ and a nonlinear modulo
operator, ${\rm{MOD}}(\bullet)$ \cite[P.127]{Fischer2002}. The output signal of THP is equivalent to the following equation \cite{JSAC_Xing2012}
\begin{align}
{\bf{x}}_0=({\bf{I}}+{\bf{B}})^{-1}(\underbrace{{\bf{a}}+{\bf{d}}}_{\triangleq {\bf{s}}}),
\end{align} where the vector ${\bf{d}}$ guarantees ${\bf{x}}_0$ in a finite region and it can be simply removed at receiver by modulo operation  \cite[P.127]{Fischer2002}.
When the elements of ${\bf{a}}$ are independent and identically distributed (i.i.d.) over the constellation and the dimension of modulation constellation is large, ${\bf{x}}_0$ can also be considered as i.i.d. \cite[P.131]{Fischer2002}, i.e., ${\bf{R}}_{{\bf{x}}_0} =\sigma_s^2{\bf{I}}_N$. For high dimensional modulation constellations, it also holds that $\sigma_s^2\approx\sigma_a^2$ irrespective of a scaling factor as the scalar factor is almost equivalent to one \cite[P.134]{Fischer2002}. In this case, the received signal at the destination is
\begin{align}
{\bf{r}} = [{\prod_{k=1}^K}{\bf{H}}_{k}{\bf{P}}_k]{\bf{x}}_0  + \sum_{k=1}^{K-1}\{ [\prod_{l={k+1}}^K{\bf{H}}_{l}{\bf{P}}_l]{\bf{n}}_k
\}+{\bf{n}}_K.
\end{align}With THP the data detection MSE matrix at the destination is
\begin{align}
\label{MSE_Matrix_3}
{\boldsymbol {\Phi}}_{\rm{MSE}}({\bf{G}},\{{\bf{P}}_k\}_{k=1}^K,{\bf{B}})=
\mathbb{E}\{({\bf{G}}{\bf{r}}-{\bf{s}})({\bf{G}}{\bf{r}}-{\bf{s}})^{\rm{H}}\}.
\end{align}

With the definition of an auxiliary matrix
\begin{align}
{\bf{C}}={\bf{I}}+{\bf{B}}
\end{align}
the previous MSE matrices given by (\ref{MSE_Matrix_1}), (\ref{MSE_Matrix_2}) and (\ref{MSE_Matrix_3}) are unified into the following formulation
\begin{align}
\label{MSE_final}
&{\boldsymbol {\Phi}}_{\rm{MSE}}({\bf{G}},\{{\bf{P}}_k\}_{k=1}^K,{\bf{C}})\nonumber \\
=&{\bf{G}}[{\bf{
H}}_{K}{\bf{P}}_K{\bf{R}}_{{\bf{x}}_{K-1}}{\bf{P}}_K^{\rm{H}} {\bf{
H}}_{K}^{\rm{H}}+{\bf{R}}_{n_K}]{\bf{G}}^{\rm{H}}+ {\bf{C}}{\bf{C}}^{\rm{H}}\sigma^2_{{\bf{a}}} \nonumber \\
& -\sigma^2_{{\bf{a}}}{\bf{C}}[\prod_{k=1}^K{\bf{H}}_{k}{\bf{P}}_k]^{\rm{H}}{\bf{G}}^{\rm{H}}-
{\bf{G}}[\prod_{k=1}^K{\bf{ H}}_{k}{\bf{P}}_k]{\bf{C}}^{\rm{H}}\sigma^2_{{\bf{a}}}.
\end{align}

\section{Problem Formulation}
\label{Problem_Formulation}

The considered optimization problem of transceiver designs aims at minimizing a matrix-monotone increasing function of the MSE matrix \cite{XingTSP2013}. For example, regarding linear transceiver designs, a series of performance metrics but not all can be formulated as additively Schur-convex/Schur-concave functions of the diagonal elements of the MSE matrix ${\boldsymbol {\Phi}}_{\rm{MSE}}({\bf{G}},\{{\bf{P}}_k\}_{k=1}^K,{\bf{C}})$, i.e., ${\textbf{d}}({\boldsymbol {\Phi}}_{\rm{MSE}}({\bf{G}},\{{\bf{P}}_k\}_{k=1}^K,{\bf{C}}))$ where symbol ${\bf{d}}({\bf{Z}})$ denotes a vector consisting of the diagonal elements of ${\bf{Z}}$, i.e., ${\bf{d}}({\bf{Z}})=[[{\bf{Z}}]_{1,1},[{\bf{Z}}]_{2,2},\cdots,[{\bf{Z}}]_{N,N}
]^{\rm{T}}$. In the following, we will discuss the considered transceiver designs case by case. Some related fundamentals of majorization theory is given in Appendix~\ref{Appendix_Majorization}.

(1) \textit{Weighted MSE}: With the data MSE matrix defined in (\ref{MSE_final}), weighted MSE can be directly written as
\begin{align}
\label{Obj 1}
\text{Obj. 1:} \ \ {\rm{Tr}}[{\bf{W}}{\boldsymbol { \Phi}}_{\rm{MSE}}({\bf{G}},\{{\bf{P}}_k\}_{k=1}^K,{\bf{C}}={\bf{I}})]
\end{align} where the weighting matrix ${\bf{W}}$ is a positive semi-definite matrix. This is different from the work in \cite{Sampth01} which only restricts to diagonal weighting matrices.

(2) \textit{Capacity}: Capacity maximization is another important and widely used performance metric for transceiver design. The capacity maximization is equivalent to minimize the following objective function
\begin{align}
\label{Obj 2}
\text{Obj. 2:}\ {\rm{log}}|{\boldsymbol {\Phi}}_{\rm{MSE}}({\bf{G}},\{{\bf{P}}_k\}_{k=1}^K,{\bf{C}}={\bf{I}})|.
\end{align}

(3) \textit{Additively Schur-convex}: design. In general, when a certain fairness in the sense of arithmetic mean is required such as worst/MAX MSE minimization, the objective function can be represented as \cite{Palomar03}
\begin{align}
\label{Obj 3}
\text{Obj. 3:}\ &f_{\rm{A-Schur}}^{\rm{Convex}}[{\bf{d}}({\boldsymbol \Phi}_{\rm{MSE}}({\bf{G}},\{{\bf{P}}_k\}_{k=1}^K,{\bf{C}}={\bf{I}}))]
\end{align}where $f_{\rm{A-Schur}}^{\rm{Convex}}(\bullet)$ is an increasing additively Schur-convex (A-Schur-Convex) function. Based on Lemma 1 in Appendix~\ref{Appendix_Majorization}, we can justify whether a function is A-Schur-Convex.

(4) \textit{Additively Schur-concave}: When a preference is given to certain data streams (e.g., the data streams with better channel state information are more preferred), the objective function can be written as \cite{Palomar03}
\begin{align}
\label{Obj 4}
\text{Obj. 4:}\ &f_{\rm{A-Schur}}^{\rm{Concave}}[{\bf{d}}({\boldsymbol \Phi}_{\rm{MSE}}({\bf{G}},\{{\bf{P}}_k\}_{k=1}^K,{\bf{C}}={\bf{I}}))]
\end{align}
where $f_{\rm{A-Schur}}^{\rm{Concave}}(\bullet)$ is an increasing additively Schur-concave (A-Schur-Concave) function. For example, weighted MSE minimization with diagonal weighting matrices is a special case of this kind of objective functions. Using Lemma 1 in Appendix~\ref{Appendix_Majorization}, we can justify whether a function is A-Schur-Concave or not.

For nonlinear transceiver designs, a series of performance metrics can be formulated as multiplicatively Schur-convex/Schur-concave functions of the diagonal elements of ${\boldsymbol {\Phi}}_{\rm{MSE}}({\bf{G}},\{{\bf{P}}_k\}_{k=1}^K,{\bf{C}})$.

(5) \textit{Multiplicatively Schur-convex}: With a certain fairness requirement is added on the geometric mean of the transmitted data streams, the objective function can be written as \cite[P.463]{Palomar2007}
\begin{align}
\label{Obj 5}
\text{Obj. 5:}\ &f_{\rm{M-Schur}}^{\rm{Convex}}[{\bf{d}}({\boldsymbol \Phi}_{\rm{MSE}}({\bf{G}},\{{\bf{P}}_k\}_{k=1}^K,{\bf{C}}))]
\end{align}
where $f_{\rm{M-Schur}}^{\rm{Convex}}(\bullet)$ is an increasing multiplicatively Schur-convex  (M-Schur-Convex) function. Based on Lemma 2 in Appendix~\ref{Appendix_Majorization}, we can justify whether a function is M-Schur-Convex.

(6) \textit{Multiplicatively Schur-concave}: With THP or DFE structure, when some preference is added to different data stream via using different weighting factors, the objective function can be written as \cite[P.466]{Palomar2007}
\begin{align}
\label{Obj 6}
\text{Obj. 6:}\ &f_{\rm{M-Schur}}^{\rm{Concave}}[{\bf{d}}({\boldsymbol \Phi}_{\rm{MSE}}({\bf{G}},\{{\bf{P}}_k\}_{k=1}^K,{\bf{C}}))]
\end{align}
where $f_{\rm{M-Schur}}^{\rm{Concave}}(\bullet)$ is an increasing multiplicatively Schur-concave (M-Schur-Concave) function. Using Lemma 2 in Appendix~\ref{Appendix_Majorization}, we can justify whether a function is M-Schur-Concave or not.

In summary, the optimization problem of transceiver designs can be formulated as follows
\begin{align}\label{original_optimization}
 & \min_{{\bf{G}},\{,{\bf{P}}_k\},{\bf{C}}} \ \  {f}\left[{\boldsymbol \Phi}_{\rm{MSE}}({\bf{G}},\{{\bf{P}}_k\}_{k=1}^K,{\bf{C}})\right] \nonumber \\
& \ \ \ \ {\rm{s.t.}} \ \ \ \
\ \ {\rm{Tr}}({\bf{P}}_k{\bf{R}}_{{\bf{x}}_{k-1}}{\bf{P}}_k^{\rm{H}})\le P_k \nonumber \\
&\ \ \ \ \ \ \ \ \ \ \ \ \ \ {\bf{P}}_k{\bf{R}}_{{\bf{x}}_{k-1}}{\bf{P}}_k^{\rm{H}}  \preceq {\bf{R}}_{{\bf{s}}_k} \nonumber \\
& \ \ \ \ \ \ \ \ \ \ \ \ \ \ [{\bf{C}}]_{i,i}=1, \ \ [{\bf{C}}]_{i,j}=0 \ \ {\text{for}} \ \ i>j
\end{align}where ${f}(\bullet)$ is a matrix-monotone increasing function.
The final two constraints come from the fact that ${\bf{B}}$ is a strictly lower triangular matrix.

It is obvious that there is no constraint on the equalizer ${\bf{G}}$. Thus for the optimal equalizer we can simply differentiate the trace of (\ref{MSE_final}) with respect to ${\bf{G}}$ and then obtain the linear minimum mean square error (LMMSE) equalizer \cite[P.344]{Kay93}
\begin{align}
\label{G} &{\bf{G}}_{\rm{LMMSE}}\nonumber \\
& =\sigma_{{\bf{a}}}^2{\bf{C}}[{\prod}_{k=1}^K{\bf{ H}}_{k}{\bf{P}}_k]^{\rm{H}}[{\bf{
H}}_{K}{\bf{P}}_K{\bf{R}}_{{\bf{x}}_{K-1}}{\bf{P}}_K^{\rm{H}} {\bf{
H}}_{K}^{\rm{H}}+{\bf{R}}_{n_K}]^{-1},
\end{align}which has the following property \cite{Palomar03}
\begin{align}
\label{G_OPT}
{\boldsymbol \Phi}_{\rm{MSE}}({\bf{G}}_{\rm{LMMSE}},\{{\bf{P}}_k\}_{k=1}^K,{\bf{C}}) \preceq {\boldsymbol \Phi}_{\rm{MSE}}({\bf{G}},\{{\bf{P}}_k\}_{k=1}^K,{\bf{C}}).
\end{align} Because ${f}(\bullet)$ is  matrix-monotone increasing function, (\ref{G_OPT}) implies that ${\bf{G}}_{\rm{LMMSE}}$ minimizes the objective function in (\ref{original_optimization}). Plugging the optimal equalizer of (\ref{G}) into (\ref{MSE_final}), we directly have
\begin{align}
\label{MSE_Matrix_Orig}
{\boldsymbol {\Phi}}_{\rm{MSE}}({\bf{G}}_{\rm{LMMSE}},\{{\bf{P}}_k\}_{k=1}^K,{\bf{C}})={\bf{C}}{\boldsymbol \Phi}_{\rm{LMMSE}}(\{{\bf{P}}_k\}_{k=1}^K){\bf{C}}^{\rm{H}}
\end{align}where the inner term on the righthand side is just the MSE matrix derived for linear transceivers
\begin{align}
&{\boldsymbol \Phi}_{\rm{LMMSE}}(\{{\bf{P}}_k\}_{k=1}^K) =\sigma_{{\bf{a}}}^2{\bf{I}}_N-\sigma_{{\bf{a}}}^2[\prod_{k=1}^K{\bf{ H}}_{k}{\bf{P}}_k]^{\rm{H}}\nonumber \\
&\times[{\bf{
H}}_{K}{\bf{P}}_K{\bf{R}}_{{\bf{x}}_{K-1}}{\bf{P}}_K^{\rm{H}} {\bf{
H}}_{K}^{\rm{H}}+{\bf{R}}_{n_K}]^{-1}[\prod_{k=1}^K{\bf{ H}}_{k}{\bf{P}}_k]\sigma_{{\bf{a}}}^2.
\end{align}

\begin{figure*}[!ht]
\centering
\includegraphics[width=.6\textwidth]{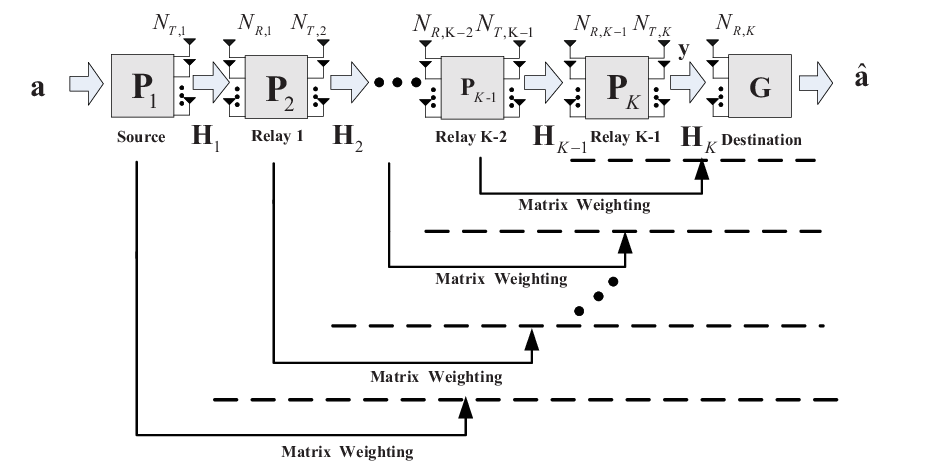}
\caption{The understanding of multi-hop AF MIMO relaying from the viewpoint of matrix weighting version operations.}\label{fig:4}
\end{figure*}

For multi-hop AF MIMO relaying systems, the received signal at the $k^{\rm{th}}$ relay node depends on the forwarding matrices at all preceding relays, and it makes the power allocations at different relays couple with each other. In order to simplify the problem substantially, we first define the following new variable in terms of ${\bf{P}}_k$:
\begin{align}
\label{F_definition}
{\bf{F}}_k&\triangleq
{\bf{P}}_k{\bf{R}}_{{\bf{n}}_{k-1}}^{1/2}\nonumber \\
& \ \ \ \  \times (
\underbrace{{\bf{R}}_{{\bf{n}}_{k-1}}^{-1/2}{\bf{
H}}_{k-1}{\bf{F}}_{k-1}{\bf{F}}_{k-1}^{\rm{H}}{\bf{
H}}_{k-1}^{\rm{H}}{\bf{R}}_{{\bf{n}}_{k-1}}^{-1/2}+{\bf{I}}}_{\triangleq {\boldsymbol \Pi}_{k-1}})^{1/2}{\bf{Q}}_{k}^{\rm{H}},
\end{align}where ${\bf{Q}}_k$ is an unknown unitary matrix.
The introduction of ${\bf{Q}}_k$ comes from the fact
that for a positive semi-definite matrix ${\bf{M}}$, its square roots generally has the form ${\bf{M}}^{1/2}{\bf{Q}}$ where ${\bf{Q}}$ is a unitary matrix. Note that at the source node  ${\bf{F}}_1=\sigma_{{\bf{a}}}{\bf{P}}_1{\bf{Q}}_1^{\rm{H}}$. Meanwhile, with the new variables ${\bf{F}}_k$,  the corresponding transmit covariance matrix at the $k^{\rm{th}}$ node can be rewritten as
\begin{align}
\label{constraint}
{\bf{P}}_{k}{\bf{R}}_{{\bf{x}}_{k-1}}{\bf{P}}_{k}^{\rm{H}}={\bf{F}}_k{\bf{F}}_k^{\rm{H}}
\end{align}based on which there will no coupled variables in the constraints.

With the new definition in (\ref{F_definition}), the matrix ${\boldsymbol { \Phi}}_{\rm{LMMSE}}(\{{\bf{Q}}_k\},\{{\bf{F}_k}\})$ is transformed to be a more compact formulation
\begin{align}
\label{MSE_Matrix_M}
&{\boldsymbol { \Phi}}_{\rm{LMMSE}}(\{{\bf{Q}}_k\},\{{\bf{F}_k}\})\nonumber \\
=&\sigma_{{\bf{a}}}^2{\bf{I}}_N-\sigma_{{\bf{a}}}^2[\prod_{k=1}^K{\boldsymbol \Pi}_k^{-1/2}{\bf{R}}_{{\bf{n}}_k}^{-1/2}{\bf{\bar H}}_k{\bf{F}}_k{\bf{Q}}_k]^{\rm{H}}\nonumber \\
& \ \ \ \ \ \ \ \ \ \ \ \ \times[\prod_{k=1}^K\underbrace{{\boldsymbol \Pi}_k^{-1/2}{\bf{R}}_{{\bf{n}}_k}^{-1/2}{\bf{H}}_k{\bf{F}}_k}_{\triangleq  {\boldsymbol A}_{k}}{\bf{Q}}_k]\nonumber \\
 =&\sigma_{{\bf{a}}}^2{\bf{I}}_N-\sigma_{{\bf{a}}}^2{\bf{Q}}_1^{\rm{H}}{\boldsymbol A}_{1}^{\rm{H}}{\bf{Q}}_2^{\rm{H}}\cdots{\boldsymbol A}_{K}^{\rm{H}}{\boldsymbol A}_{K}\cdots{\bf{Q}}_2{\boldsymbol A}_{1}{\bf{Q}}_1.
\end{align}
It is obvious that with the new variables ${\bf{F}}_k$'s, the constraints become independent with each other. Putting (\ref{constraint}) and (\ref{MSE_Matrix_M}) into the original optimization (\ref{original_optimization}), the transceiver design problem can be reformulated as
\begin{align}
\label{Optimziation_Original}
 & \min_{\{{\bf{F}}_k\},\{{\bf{Q}}_k\},{\bf{C}}} \ \  {f}\left[{\bf{C}}{\boldsymbol \Phi}_{\rm{LMMSE}}(\{{\bf{F}}_k\},\{{\bf{Q}}_k\}){\bf{C}}^{\rm{H}}\right] \nonumber \\
& \ \ \ \ \ {\rm{s.t.}} \ \ \ \ \ \
\ \ {\rm{Tr}}({\bf{F}}_k{\bf{F}}_k^{\rm{H}})\le P_k \nonumber \\
&\ \ \ \ \ \ \ \ \ \ \ \ \ \ \ \ \ {\bf{F}}_k{\bf{F}}_k^{\rm{H}}  \preceq {\bf{R}}_{{\bf{s}}_k} \nonumber \\
& \ \ \ \ \ \ \ \ \ \ \ \ \ \ \ \ \ [{\bf{C}}]_{i,i}=1 \nonumber \\
& \ \ \ \ \ \ \ \ \ \ \ \ \ \ \ \ \ [{\bf{C}}]_{i,j}=0 \ \ {\text{for}} \ \ i>j.
\end{align}

\noindent \textbf{Matrix Version Weighting Operation Interpretation:}

If the following analysis, we will simply set $\sigma_{\bf{a}}^2=1$ without loss of generality.
It is worth noting that based on the definition of ${\bf{A}}_k$ in (\ref{MSE_Matrix_M}) it can be proved that ${\bf{I}}-{\bf{Q}}_{k}^{\rm{H}}{\boldsymbol A}_{k}^{\rm{H}} {\boldsymbol A}_{k}{\bf{Q}}_{k}=({\bf{Q}}_k^{\rm{H}}{\bf{F}}_k^{\rm{H}}{\bf{H}}_k^{\rm{H}}{\bf{R}}_{{\bf{n}}_k}^{-1}
{\bf{H}}_k{\bf{F}}_k{\bf{Q}}_k+{\bf{I}})^{-1}$. Taking ${\bf{F}}_k{\bf{Q}}_k$ as a precoding matrix, ${\bf{I}}-{\bf{Q}}_{k}^{\rm{H}}{\boldsymbol A}_{k}^{\rm{H}} {\boldsymbol A}_{k}{\bf{Q}}_{k}$ is the data detection MSE matrix for LMMSE estimator in the $k^{\rm{th}}$ hop \cite[Euq.3.21]{Palomar2007}. For LMMSE estimators the estimated signal is independent of the residual noise \cite{Kay93} and then ${\bf{Q}}_{k}^{\rm{H}}{\boldsymbol A}_{k}^{\rm{H}} {\boldsymbol A}_{k}{\bf{Q}}_{k}$ is the covariance matrix of the estimated signal at the $k^{\rm{th}}$ relay node. Roughly speaking, the singular values of ${\boldsymbol A}_{k}{\bf{Q}}_{k}$ reflect the strength of the recovered signals.

Just as discussed in \cite{XingCommL2013}, AF MIMO relaying can be recognized as a certain matrix version weighting operation. For example, if we only focus the final two hops, it is a standard dual hop AF MIMO relaying system and its MSE matrix can be written in the following form
\begin{align}
&\underbrace{{\bf{Q}}_{K-1}^{\rm{H}}{\boldsymbol A}_{K-1}^{\rm{H}}}_{{\boldsymbol W}^{\rm{H}}}({\bf{I}}-{\bf{Q}}_{K}^{\rm{H}}{\boldsymbol A}_{K}^{\rm{H}} {\boldsymbol A}_{K}{\bf{Q}}_{K})\underbrace{{\boldsymbol A}_{K-1}{\bf{Q}}_{K-1}}_{{\boldsymbol W}}\nonumber \\
&+\underbrace{{\bf{I}}-{\bf{Q}}_{K-1}^{\rm{H}}{\boldsymbol A}_{K-1}^{\rm{H}}{\boldsymbol A}_{K-1}{\bf{Q}}_{K-1}}_{{\boldsymbol \Pi}}.
\end{align}In accordance to the definition of matrix version weighting in \cite{XingCommL2013}, the matrix weighting of  the $(K-1)^{\rm{th}}$ hop on the $K^{\rm{th}}$ hop is carried out by multiplying a matrix version slope ${\boldsymbol{W}}$ and adding a matrix version intercept ${\boldsymbol \Pi}$. Notice that the matrix version intercept is just the MSE matrix for the $K-1$ hop. As shown in Fig.~\ref{fig:4}, repeat this process and finally we will have the exact formula of ${\boldsymbol \Phi}_{\rm{LMMSE}}(\{{\bf{F}}_k\},\{{\bf{Q}}_k\})$. Interestingly, the roles of THP and DFE also fall into the category of this kind of matrix weighting operations with  matrix version slope ${\boldsymbol W}={\bf{C}}^{\rm{H}}$ and matrix version intercept ${\boldsymbol \Pi}={\bf{0}}$. In this case, the matrix version intercept equals zero because THP or DFE does not introduce noises.

\section{Reformulation of the Considered Optimization Problem}
\label{Simplified_Optimization}

In the optimization problem (\ref{Optimziation_Original}) discussed above, there are three kinds of variables, i.e., ${\bf{C}}$, ${\bf{Q}}_k$'s and ${\bf{F}}_k$'s. In this section, we will try our best to simplify the problem (\ref{Optimziation_Original}) by first deriving the optimal solutions of ${\bf{C}}$ and ${\bf{Q}}_k$'s to be the functions of ${\bf{F}}_k$'s and then the number of variables will be significantly reduced.
\subsection{Optimal ${\bf{C}}$}
Different from the optimal solution of the equalizer, the optimal solutions of ${\bf{C}}$ are different for linear and nonlinear transceivers. For linear transceivers, ${\bf{C}}$ is a constant identity matrix.
On the other hand, for nonlinear transceiver designs with DFE or THP, we have the following result \cite{JSAC_Xing2012}\begin{align}
\label{C_opt}
{\bf{C}}_{\rm{opt}}={\rm{diag}}\{[{\bf{L}}_{1,1}, \cdots, {\bf{L}}_{N,N}]^{\rm{T}}\}{\bf{L}}^{-1},
\end{align}where the lower triangular matrix ${\bf{L}}$ is defined based on the Cholesky factorization of ${\boldsymbol \Phi}_{\rm{LMMSE}}(\{{\bf{F}}_k\},\{{\bf{Q}}_k\})$, i.e., ${\boldsymbol \Phi}_{\rm{LMMSE}}(\{{\bf{F}}_k\},\{{\bf{Q}}_k\})
={\bf{L}}{\bf{L}}^{\rm{H}}$.
Based on (\ref{C_opt}) at the optimum values the objective functions in Cases 5 and 6 are equivalent to
\begin{align}
\text{Obj. 5:}\ &f_{\rm{M-Schur}}^{\rm{Convex}}({\bf{d}}^2[{\bf{L}}]) , \ \ {\boldsymbol \Phi}_{\rm{LMMSE}}(\{{\bf{F}}_k\},\{{\bf{Q}}_k\})={\bf{L}}{\bf{L}}^{\rm{H}}, \\
\text{Obj. 6:}\ &f_{\rm{M-Schur}}^{\rm{Concave}}({\bf{d}}^2[{\bf{L}}]) , \ \ {\boldsymbol \Phi}_{\rm{LMMSE}}(\{{\bf{F}}_k\},\{{\bf{Q}}_k\})={\bf{L}}{\bf{L}}^{\rm{H}}.
\end{align}

\subsection{Optimal ${\bf{Q}}_k$'s}

The derivation of optimal ${\bf{Q}}_k$'s is based on matrix inequality theory especially majorization theory. Defining a unitary matrix ${\bf{U}}_{\boldsymbol{\Theta}}$ based on the eigenvalue decomposition (EVD) of
\begin{align}
{\boldsymbol A}_{1}^{\rm{H}}{\bf{Q}}_2^{\rm{H}}{\boldsymbol A}_{2}^{\rm{H}}\cdots{\boldsymbol A}_{K}^{\rm{H}}{\boldsymbol A}_{K}\cdots{\boldsymbol A}_{2}{\bf{Q}}_2{\boldsymbol A}_{1}={\bf{U}}_{{\boldsymbol \Theta}}{\boldsymbol{\Lambda}}_{{\boldsymbol \Theta}}{\bf{U}}_{{\boldsymbol \Theta}}^{\rm{H}}
\end{align}
with eigenvalues in decreasing order, following the same logic in \cite{JSAC_Xing2012,XingTSP2013} it can be proved that the optimal ${\bf{Q}}_1$ equals
\begin{align}
\label{Q_0_opt}
 {\bf{Q}}_1={\bf{U}}_{{\boldsymbol \Theta}}{\bf{U}}_{\boldsymbol{\Omega}}^{\rm{H}}
\end{align} in which ${\bf{U}}_{\boldsymbol{\Omega}}$ has the following solution
\begin{align}
\label{U}
{\bf{U}}_{\boldsymbol{\Omega}}=\left\{ {\begin{array}{*{20}c}
   {{\bf{U}}_{\bf{W}} \ \ \ \text{for Obj 1}}  \\
   {{\bf{U}}_{\rm{Arb}}\ \ \text{for Obj 2}}  \\
   {{\bf{Q}}_{\rm{DFT}} \ \ \text{for Obj 3}}  \\
   { \ \ {\bf{I}}_N \ \ \ \  \text{for Obj 4}}  \\
   { \ {\bf{Q}}_{\bf{T}} \ \ \ \  \text{for Obj 5}}  \\
   { \ \ {\bf{I}}_N \ \ \ \  \text{for Obj 6}}  \\
\end{array}} \right.
\end{align}where ${\bf{U}}_{\bf{W}}$ is the unitary matrix of the EVD of ${\bf{W}}$ with eigenvalues in decreasing order and ${\bf{U}}_{\rm{Arb}}$ is an arbitrary unitary matrix with proper dimensionality. Moreover, the unitary matrix $ {\bf{Q}}_{\rm{DFT}}$ is discrete Fourier transform (DFT) matrix and the matrix ${\bf{Q}}_{\bf{T}}$ is the unitary matrix that makes the Cholesky factorization matrix of ${\bf{Q}}_1^{\rm{H}}({\bf{I}}-{\boldsymbol A}_{1}^{\rm{H}}{\bf{Q}}_2^{\rm{H}}\cdots{\boldsymbol A}_{K}^{\rm{H}}{\boldsymbol A}_{K}\cdots{\bf{Q}}_2{\boldsymbol A}_{1}){\bf{Q}}_1$ have identical diagonal elements.

\textsl{Proof:} The detailed proof can be found in Subsection A in Section IV in \cite{JSAC_Xing2012} and Appendix~A in \cite{XingTSP2013}.   $\blacksquare$

Besides ${\bf{Q}}_1$ discussed above, the optimal ${\bf{Q}}_k$'s for $k=2,\cdots,K$ should satisfy the following property \cite{XingTSP2013,JSAC_Xing2012}
\begin{align}
\label{Q_k_opt}
{\bf{Q}}_k={\bf{V}}_{{\boldsymbol{A}}_{k}}{\bf{U}}_{{\boldsymbol{A}}_{k-1}}^{\rm{H}}, \ \ k=2,\cdots,K.
\end{align}where the unitary matrices ${\bf{U}}_{{\boldsymbol{A}}_k}$ and ${\bf{V}}_{{\boldsymbol{A}}_{k}}$ come from the singular value decomposition (SVD) ${\boldsymbol{A}}_k={\bf{U}}_{{\boldsymbol{A}}_k}
{\boldsymbol{\Lambda}}_{{\boldsymbol{A}}_k}{\bf{V}}_{{\boldsymbol{A}}_k}^{\rm{H}}$ with ${\boldsymbol{\Lambda}}_{{\boldsymbol{A}}_k} \searrow$.

\textsl{Proof:} The detailed proof can be found in Appendix~E in \cite{JSAC_Xing2012} and Appendix~B in \cite{XingTSP2013}.   $\blacksquare$

\subsection{The Reformulated optimization problems}

Based on the optimal solutions of ${\bf{C}}$ and ${\bf{Q}}_k$'s listed particularly, the original optimization problem (\ref{Optimziation_Original}) becomes
\begin{align}
\label{specific_one}
&¡¡\min_{\{{\bf{F}}_k\}} \ \ \ g \left(\left\{{\boldsymbol \lambda}({\bf{F}}^{\rm{H}}_k{\bf{H}}_{k}^{\rm{H}}
{\bf{R}}_{{\bf{n}}_{k}}^{-1}{\bf{H}}_k{\bf{F}}_k)\right\}_{k=1}^K\right) \nonumber \\
&¡¡\  {\rm{s.t.}} \ \ \ \  {\rm{Tr}}({\bf{F}}_k{\bf{F}}_k^{\rm{H}})\le P_k \nonumber \\
&¡¡\ \ \ \ \ \ \ \ \ {\bf{F}}_k{\bf{F}}_{k}^{\rm{H}} \preceq {\bf{R}}_{{\bf{s}}_k}
\end{align}where ${\boldsymbol \lambda}({\bf{Z}})$ denotes the vector consisting of eigenvalues, i.e., ${\boldsymbol \lambda}({\bf{Z}})=[\lambda_1({\bf{Z}}),\lambda_2({\bf{Z}}),\cdots,
\lambda_N({\bf{Z}})]^{\rm{T}}$.
Furthermore, it should be highlighted that the objective $g(\bullet)$ is a monotonically decreasing function with respect to the following big column vector
\begin{align}
\label{Big_Vector}
[{\boldsymbol \lambda}({\bf{F}}^{\rm{H}}_1{\bf{H}}_{1}^{\rm{H}}
{\bf{R}}_{{\bf{n}}_{1}}^{-1}{\bf{H}}_1{\bf{F}}_1),\cdots,{\boldsymbol \lambda}({\bf{F}}^{\rm{H}}_K{\bf{H}}_{K}^{\rm{H}}
{\bf{R}}_{{\bf{n}}_{K}}^{-1}{\bf{H}}_K{\bf{F}}_K)]^{\rm{T}}.
\end{align}Note that for any a given performance metric, the objective function of (\ref{specific_one}) will be a specific decreasing function in the vector given by (\ref{Big_Vector}) and ${\boldsymbol \lambda}({\bf{F}}^{\rm{H}}_k{\bf{H}}_{k}^{\rm{H}}
{\bf{R}}_{{\bf{n}}_{k}}^{-1}{\bf{H}}_k{\bf{F}}_k)$'s are coupled with each other. In this paper, various transceiver designs are investigated from a unified viewpoint and then we are only concerned with the common characteristic of the transceiver designs. This is the motivation of the following conversion.

For each ${\boldsymbol \lambda}({\bf{F}}^{\rm{H}}_k{\bf{H}}_{k}^{\rm{H}}
{\bf{R}}_{{\bf{n}}_{k}}^{-1}{\bf{H}}_k{\bf{F}}_k)$, the objective function of (\ref{specific_one}) is a decreasing function. If the affects of the formula of the objective function are neglected, we will have a more general optimization problem which has the following formula \begin{align}
\label{subproblem_matrix}
& \ \max_{{\bf{F}}_k} \ \ {\boldsymbol \lambda}\left({\bf{F}}^{\rm{H}}_k{\bf{H}}_{k}^{\rm{H}}
{\bf{R}}_{{\bf{n}}_k}^{-1}{\bf{H}}_k{\bf{F}}_k\right) \nonumber \\
& \ \ {\rm{s.t.}}  \ \ \ \  {\rm{Tr}}({\bf{F}}_k{\bf{F}}_k^{\rm{H}})\le P_k \nonumber \\
&¡¡\ \ \ \ \ \ \ \   \ \   {\bf{F}}_k{\bf{F}}_{k}^{\rm{H}} \preceq {\bf{R}}_{{\bf{s}}_k}
\end{align}where the objective function is a vector function instead of a scalar function (taking each element $\lambda_i({\bf{F}}^{\rm{H}}_k{\bf{H}}_{k}^{\rm{H}}
{\bf{R}}_{{\bf{n}}_k}^{-1}{\bf{H}}_k{\bf{F}}_k)$ as a function in ${\bf{F}}_k$).
In (\ref{subproblem_matrix}), there are no more constraints introduced by the formula of the objective function. Thus for the variable ${\bf{F}}_k$ (\ref{subproblem_matrix}) is more general than (\ref{specific_one}) which has a specific objective function. In other words, for the variable ${\bf{F}}_k$ the optimal solution set of (\ref{specific_one}) is a subset of the Pareto optimal solution set of (\ref{subproblem_matrix}) \cite[P.177]{Boyd04}\footnote{When the optimal objective value has many solutions, we only choose the ones with ${\bf{F}}_k{\bf{F}}_k^{\rm{H}}$ being the largest in positive semi-definite cone}.
For all the possible $g(\bullet)$, the union set of the optimal solution set of (\ref{specific_one}) will be the Pareto optimal solution set of (\ref{subproblem_matrix}).
 The common characteristics of all the Pareto optimal solutions of (\ref{subproblem_matrix}) must also be owned by the optimal solution of  (\ref{specific_one}).
As we consider the union set of the optimal solution set of (\ref{specific_one}),
the coupling effects between ${\boldsymbol \lambda}({\bf{F}}^{\rm{H}}_k{\bf{H}}_{k}^{\rm{H}}
{\bf{R}}_{{\bf{n}}_{k}}^{-1}{\bf{H}}_k{\bf{F}}_k)$'s will disappear. The specific coupling relationship between ${\boldsymbol \lambda}({\bf{F}}^{\rm{H}}_k{\bf{H}}_{k}^{\rm{H}}
{\bf{R}}_{{\bf{n}}_{k}}^{-1}{\bf{H}}_k{\bf{F}}_k)$'s is determined by specific objective function and it can be understood as the principle how to select the exact optimal solution for a specific objective function from the union set. In the following work, we will focus on (\ref{subproblem_matrix}) to derive the common characteristics of its Pareto optimal set, which are also the characteristics of the optimal solutions of (\ref{specific_one}).

Note that in some cases for the optimization problem (\ref{subproblem_matrix}) there may be one physical limitation that the number of eigenchannels used in each hop should better not be larger than that of the data streams. This limitation may result in a rank constraint, i.e., ${\rm{Rank}}\{{\bf{F}}_k{\bf{F}}_{k}^{\rm{H}}\}\le N$. However, we discover that there is no need to consider this  physical limitation in the following analysis.  When this physical limitation is considered, it means only the first $N$ elements of ${\boldsymbol \lambda}({\bf{F}}^{\rm{H}}_k{\bf{H}}_{k}^{\rm{H}}
{\bf{R}}_{{\bf{n}}_k}^{-1}{\bf{H}}_k{\bf{F}}_k)$ need to be considered and the optimization problem will become
\begin{align}
\label{Matrix_Rank}
& \ \max_{{\bf{F}}_k} \ \ [{\boldsymbol \lambda}\left({\bf{F}}^{\rm{H}}_k{\bf{H}}_{k}^{\rm{H}}
{\bf{R}}_{{\bf{n}}_k}^{-1}{\bf{H}}_k{\bf{F}}_k\right)]_{1:N} \nonumber \\
& \ \ {\rm{s.t.}}  \ \ \ \  {\rm{Tr}}({\bf{F}}_k{\bf{F}}_k^{\rm{H}})\le P_k\nonumber \\
 & \ \  \ \ \ \ \ \ \ \ {\bf{F}}_k{\bf{F}}_{k}^{\rm{H}} \preceq {\bf{R}}_{{\bf{s}}_k}
\end{align}whose Pareto optimal solution set is also a subset of the Pareto optimal solution set of (\ref{subproblem_matrix}).
Following the previous logic, all the common characteristics of the Pareto optimal solutions of (\ref{subproblem_matrix}) will be inherited by the Pareto optimal solutions of (\ref{Matrix_Rank}) as well. As (\ref{subproblem_matrix}) is easier to analyze than (\ref{Matrix_Rank}), in the following we only focus on  (\ref{subproblem_matrix}).

%

\section{Transceiver Designs with Pure Shaping Constraints}
\label{Pure_Shaping_Constraint}

In the optimization problem (\ref{subproblem_matrix}), except the rank constraint, there are still two constraints i.e., sum power constraint and shaping constraint. If shaping constraint is stricter than sum power constraint, i.e., ${\rm{Tr}}({\bf{R}}_{\bf{s}_s})\le P_k$, the sum power constraint can be removed directly. Therefore, the optimization problem (\ref{subproblem_matrix}) can be written as
\begin{align}
& \ \max_{{\bf{F}}_k} \ \ {\boldsymbol \lambda}\left({\bf{F}}^{\rm{H}}_k{\bf{H}}_{k}^{\rm{H}}
{\bf{R}}_{{\bf{n}}_k}^{-1}{\bf{H}}_k{\bf{F}}_k\right) \nonumber \\
& \ \ {\rm{s.t.}}  \ \ \ {\bf{F}}_k{\bf{F}}_{k}^{\rm{H}} \preceq {\bf{R}}_{{\bf{s}}_k}.
\end{align}As previously discussed, this problem corresponds to the transceiver designs under independent power constraints per-antenna, spectral mask constraint or power constraint along a spatial direction \cite{Palomar2004}. It is obvious that the above optimization problem is equivalent to the following one
\begin{align}
\label{covariance_optimization}
&¡¡\max_{{\bf{F}}_k} \ \ \ {\boldsymbol \lambda}\left({\bf{R}}_{{\bf{n}}_k}^{-1/2}{\bf{H}}_k{\bf{F}}_k{\bf{F}}^{\rm{H}}_k{\bf{H}}_{k}^{\rm{H}}
{\bf{R}}_{{\bf{n}}_k}^{-1/2}\right) \nonumber \\
& \ \ {\rm{s.t.}}  \ \ \   {\bf{F}}_k{\bf{F}}_{k}^{\rm{H}} \preceq {\bf{R}}_{{\bf{s}}_k} \nonumber \\
& \ \ \ \ \ \ \ \ \ {\rm{Rank}}\{{\bf{F}}_k{\bf{F}}_{k}^{\rm{H}}\}\le N_{k}=\min\{N_{T,k},N_{R,k-1}\},
\end{align}where the final constraint comes from that the rank of ${\bf{F}}_k{\bf{F}}_{k}^{\rm{H}}$ is no more than the minimum number of the row and column of ${\bf{F}}_{k}$ and note that $N_{R,0}\triangleq N$. The introduction of the final constraint guarantees the equivalence between taking ${\bf{F}}_k$ as a variable and ${\bf{F}}_k{\bf{F}}_k^{\rm{H}}$ as a variable. If this constraint on the rank is neglected  there will be a rank relaxation, e.g., semi-define relaxation (SDR) \cite{Nassab2008,Huang2010,Timotheou2014}. Proving Rank relaxation is tight is usually a necessary but challenging task. Otherwise, when the relaxation is not tight the solution will be an ad-hoc one that may be even infeasible. In our work, the rank constraint is explicitly added and therefore there is no rank relaxation.

Note that ${\boldsymbol \lambda}({\bf{R}}_{{\bf{n}}_k}^{-1/2}{\bf{H}}_k{\bf{F}}_k{\bf{F}}^{\rm{H}}_k{\bf{H}}_{k}^{\rm{H}}
{\bf{R}}_{{\bf{n}}_k}^{-1/2})$ is monotonic with respect to ${\bf{F}}_k{\bf{F}}^{\rm{H}}_k$.
If the rank constraint in (\ref{covariance_optimization}) is inactive i.e., ${\rm{Rank}}\{ {\bf{R}}_{{\bf{s}}_k}\} \le N_k$,
the optimal ${\bf{F}}_k$ satisfies ${\bf{F}}_k{\bf{F}}_k^{\rm{H}}={\bf{R}}_{{\bf{s}}_k}$ and it equals
\begin{align}
{\bf{F}}_{k,{\rm{opt}}}={\bf{U}}_{{\bf{R}}_{{\bf{s}}_k}}
\left[
   {{\boldsymbol \Lambda}_{{\bf{R}}_{{\bf{s}}_k}}^{1/2}} \
   {{\bf{0}}}   \right]{\bf{U}}_{{\rm{Arb}},k}^{\rm{H}}
\end{align}where the unitary matrix ${\bf{U}}_{{\bf{R}}_{{\bf{s}}_k}} $ and the diagonal matrix ${\boldsymbol \Lambda}_{{\bf{R}}_{{\bf{s}}_k}}$ are defined based on the EVD ${\bf{R}}_{{\bf{s}}_k}={\bf{U}}_{{\bf{R}}_{{\bf{s}}_k}}{\boldsymbol \Lambda}_{{\bf{R}}_{{\bf{s}}_k}}
{\bf{U}}_{{\bf{R}}_{{\bf{s}}_k}}^{\rm{H}}$ with ${\boldsymbol{\Lambda}}_{{\bf{R}}_{{\bf{s}}_k}} \searrow$. The unitary matrix ${\bf{U}}_{{\rm{Arb}},k}$ is an arbitrary unitary matrix with proper dimensionality.
On the other hand, when ${\rm{Rank}}\{ {\bf{R}}_{{\bf{s}}_k}\} > N_k$ the optimization problem becomes more complicated. To the best of our knowledge, even for point-to-point MIMO systems, the optimal solutions in this case is largely open \cite{Palomar2004}. There are infinite matrices satisfying the rank constraints but they cannot be ordered according to positive semidefinite cone and exhaustive evaluation seems the only way to find the optimal solutions that are dependent on the channel matrices and objective functions \cite{Palomar2004}.

To avoid exhaustive evaluation, a reasonable and effective logic is to derive a lower bound of the objective function of (\ref{covariance_optimization}), based on which closed-form solutions can be derived. The inequality ${\bf{F}}_k{\bf{F}}^{\rm{H}}_k \preceq {\bf{R}}_{{\bf{s}}_k}$ can be rewritten as ${\bf{R}}_{{\bf{s}}_k}={\bf{F}}_k{\bf{F}}^{\rm{H}}_k +{\bf{R}}_{\Delta_k}$ and ${\bf{R}}_{\Delta_k}$ is positive semidefinite. Then the problem becomes to minimize ${\bf{R}}_{\Delta_k}$.
In our work, to derive the lower bound of the objective function we try to minimize the sum of the eigenvalues of ${\bf{R}}_{\Delta_k}$ i.e., ${\rm{Tr}}({\bf{R}}_{\Delta_k})$ instead of ${\bf{R}}_{\Delta_k}$. It is equivalent to maximizing the sum of the eigenvalues of ${\bf{F}}_k{\bf{F}}^{\rm{H}}_k$. Notice that when ${\bf{A}} \preceq {\bf{B}}$ it can be concluded that ${\lambda}_j({\bf{A}})\le {\lambda}_j({\bf{B}})$. Therefore, as ${\bf{F}}_k{\bf{F}}^{\rm{H}}_k \preceq {\bf{R}}_{{\bf{s}}_k}$ for the maximum trace the eigenvalues of ${\bf{F}}_k{\bf{F}}^{\rm{H}}_k$ should be
\begin{align}
{\lambda}_{j}({\bf{F}}_k{\bf{F}}^{\rm{H}}_k)=\left\{ {\begin{array}{*{20}c}
   {{\lambda}_{j}({\bf{R}}_{{\bf{s}}_k})
\ \ \ j=1:N_k}  \\
   {\ \ \ 0 \ \ \ \ \ \ \ \ {\rm{Otherwise}}}  \\
\end{array}} \right.
\end{align}based on which the shaping constraint in (\ref{covariance_optimization}) becomes to be
\begin{align}
{\bf{F}}_k{\bf{F}}_{k}^{\rm{H}}&=[{\bf{U}}_{{\bf{F}}_k}]_{:,1:N_k}[{\boldsymbol \Lambda}_{{\bf{F}}_k}]_{1:N_k,1:N_k}[{\boldsymbol \Lambda}_{{\bf{F}}_k}]_{1:N_k,1:N_k}^{\rm{T}}
[{\bf{U}}_{{\bf{F}}_k}]_{:,1:N_k}^{\rm{H}}\nonumber \\
& \preceq {\bf{R}}_{{\bf{s}}_k}
\end{align}Notice that when ${\bf{A}} \preceq {\bf{B}}$ in general we can only argue that ${\lambda}_j({\bf{A}})\le {\lambda}_j({\bf{B}})$ but we cannot say ${\bf{A}}$ and ${\bf{B}}$ have the same eigenvectors in EVD. Here, as ${\lambda}_{j}({\bf{F}}_k{\bf{F}}^{\rm{H}}_k)={\lambda}_{j}({\bf{R}}_{{\bf{s}}_k})$
for $j=1:N_k$, in Appendix~\ref{Appendix_U} it can be proved that $[{\bf{U}}_{{\bf{F}}_k}]_{:,1:N_k}=
[{\bf{U}}_{{\bf{R}}_{{\bf{s}}_k}}]_{:,1:N_k}$ and then we have
\begin{align} {\bf{F}}_k{\bf{F}}_{k}^{\rm{H}}= [{\bf{U}}_{{\bf{R}}_{{\bf{s}}_k}}]_{:,1:N_k}[{\boldsymbol \Lambda}_{{\bf{R}}_{{\bf{s}}_k}}]_{1:N_k,1:N_k}
[{\bf{U}}_{{\bf{R}}_{{\bf{s}}_k}}]_{:,1:N_k}^{\rm{H}}.
\end{align}
Finally, the proposed closed-form solution of ${\bf{F}}_k$ is
\begin{align}
\label{F_k_pure}
{\bf{F}}_{k,{\rm{opt}}}={\bf{U}}_{{\bf{R}}_{{\bf{s}}_k}}
\left[ {\begin{array}{*{20}c}
   {[{\boldsymbol \Lambda}_{{\bf{R}}_{{\bf{s}}_k}}]_{1:N_k,1:N_k}^{1/2}} & {}  \\
   {} & {{\bf{0}}}  \\
\end{array}} \right]{\bf{U}}_{{\rm{Arb}},k}^{\rm{H}}.
\end{align}

Based on the previous discussions, we discover that for the transceiver designs with pure shaping constraint, the proposed closed-form solutions of ${\bf{F}}_k$'s are independent of the objective functions and channel realizations. In other words, for the objective functions ranging from Obj. 1 to Obj. 6, ${\bf{F}}_k$'s have the same solutions. When ${\rm{Rank}}\{ {\bf{R}}_{{\bf{s}}_k}\} \le N_k$ the proposed solutions are optimal. Unfortunately when ${\rm{Rank}}\{ {\bf{R}}_{{\bf{s}}_k}\} > N_k$, the solution in (\ref{F_k_pure}) corresponds to a lower bound of the objective function of (\ref{covariance_optimization}) and the tightness of the lower bound cannot be guaranteed rigorously, but using this bound is  an effect way to avoid exhaustive evaluation among infinite candidate matrices. This conclusion is consistent with its counterpart for point-to-point MIMO systems \cite{Palomar2004}. The proposed solutions also implies that the transceiver designs with pure shaping constraint is well-suited for distributed implementation.

\noindent \textbf{Distributed Implementation}

Based on the proposed solution for ${\bf{F}}_k$ in (\ref{F_k_pure}) and its definition in (\ref{F_definition}) and together with ${\bf{R}}_{n_{k}}=\sigma_{n_k}^2{\bf{I}}_{N_{T,k}}$, it can be concluded that the forwarding matrix ${\bf{P}}_k$ at the $k^{\rm{th}}$ node equals ${\bf{P}}_{k,{\rm{opt}}}={\bf{F}}_{k,{\rm{opt}}}{\bf{Q}}_{k}
{\bf{R}}_{{\bf{x}}_{k-1}}^{-1/2}$. Shown by (\ref{F_k_pure}), computation of ${\bf{F}}_{k,{\rm{opt}}}$ does not need information exchange between neighbouring nodes. From (\ref{Q_k_opt}) it can be seen that the computation of ${\bf{Q}}_{k}$ only needs the information exchange between neighbouring nodes. The computation of ${\bf{R}}_{{\bf{x}}_{k-1}}$ is a little bit difficult. The matrix ${\bf{R}}_{{\bf{x}}_{k-1}}$ is the covariance matrix  of the received signal at the $k^{\rm{th}}$ node. Based on its definition in (\ref{R_x}) and together with (\ref{constraint}), ${\bf{R}}_{{\bf{x}}_{k-1}}$ is determined by ${\bf{F}}_{k-1,{\rm{opt}}}$ at its immediately preceding node, i.e., the $(k-1)^{\rm{th}}$ node. In other words, its computation only needs to share local information.

\section{Transceiver Designs with Joint Power Constraints}
\label{Joint_Power_Constraint}

In this section, we take a further step to investigate a more complicated case with joint power constraints. In this case, the original optimization problem (\ref{subproblem_matrix}) has the following special formulation
\begin{align}
\label{Optim_Joint_Constraint}
& \ \max_{{\bf{F}}_k} \ \ \ \ {\boldsymbol \lambda}\left({\bf{F}}^{\rm{H}}_k{\bf{H}}_{k}^{\rm{H}}
{\bf{R}}_{{\bf{n}}_k}^{-1}{\bf{H}}_k{\bf{F}}_k\right) \nonumber \\
& \ \ \ {\rm{s.t.}} \ \ \ \ {\rm{Tr}}({\bf{F}}_k{\bf{F}}_k^{\rm{H}})\le P_k \nonumber \\
&¡¡\ \ \ \ \ \ \ \ \ \ \  {\bf{F}}_k{\bf{F}}_{k}^{\rm{H}} \preceq \tau_{k,{\rm{max}}}{\bf{I}} .
\end{align}It should be highlighted that for joint power constraints, we only focus on the case in which the sum power constraint is always active. It is because if the sum power constraint is inactive, the considered optimization problem will reduce to a special case of that discussed in the previous section. Fortunately, we discover that actually the derived solution is also well-suited for the case where the sum power constraint is inactive. The formulation of joint power constraints can be interpreted as an effect way to model transceiver designs under peak power constraint \cite{Dai2012}.
The above optimization problem (\ref{Optim_Joint_Constraint}) is equivalent to the following optimization problem \cite{XingTSP2014}
\begin{align}
\label{matrix_monotonic_opt}
{\textbf{Prob. 1:}} \ & \ \max_{{\bf{F}}_k} \ \ \ \ {\bf{F}}^{\rm{H}}_k{\bf{H}}_{k}^{\rm{H}}
{\bf{R}}_{{\bf{n}}_k}^{-1}{\bf{H}}_k{\bf{F}}_k \nonumber \\
& \ \ \ {\rm{s.t.}} \ \ \ \ {\rm{Tr}}({\bf{F}}_k{\bf{F}}_k^{\rm{H}})\le P_k \nonumber \\
&¡¡\ \ \ \ \ \ \ \ \ \ \ {\bf{F}}_k{\bf{F}}_{k}^{\rm{H}} \preceq \tau_{k,{\rm{max}}}{\bf{I}}.
\end{align} It is worth noting that the optimization problem (\ref{matrix_monotonic_opt}) is in nature a multi-objective optimization problem with respect to positive semi-definite cone \cite[P.180]{Boyd04}. The role of maximizing a positive semi-definite matrix is two fold: maximizing its eigenvalues and choosing a proper  unitary matrix of its EVD \cite{XingTSP2014}. For any feasible ${\bf{F}}_k$ satisfying constraints in (\ref{matrix_monotonic_opt}), introducing any unitary matrix ${\bf{U}}$, ${\bf{U}}{\bf{F}}_k$ also satisfies the constraints and then there is no need to optimize the unitary matrix. Therefore,  (\ref{Optim_Joint_Constraint}) and  (\ref{matrix_monotonic_opt}) are equivalent. In this case, our attention is still focused on the Pareto optimal solution set. Because of the matrix version objective, directly deriving the Pareto optimal set is challenging. Necessary transformations are needed.


\subsection{The Structures of Optimal Solutions}
Defining the Pareto optimal solutions ${\bf{F}}_{k,{\rm PO}}$'s for \textbf{Prob. 1}, ${\bf{F}}_{k,{\rm PO}}$'s must satisfy all the constraints i.e., ${\rm{Tr}}({\bf{F}}_{k,{\rm{PO}}}{\bf{F}}_{k,{\rm{PO}}}^{\rm{H}})\le P_k$, ${\bf{F}}_{k,{\rm{PO}}}{\bf{F}}_{k,{\rm{PO}}}^{\rm{H}} \le \tau_{k,{\rm{max}}}{\bf{I}}$ and ${\rm{Rank}}\{{\bf{F}}_{k,{\rm{PO}}}{\bf{F}}_{k,{\rm{PO}}}^{\rm{H}} \}\le N$. Furthermore, for any given Pareto optimal solution ${\bf{F}}_{k,{\rm PO}}$, it is impossible to find a feasible ${\bf{F}}_k$ under the constraints specified in \textbf{Prob. 1} which satisfies ${\bf{F}}^{\rm{H}}_k{\bf{H}}_{k}^{\rm{H}}
{\bf{R}}_{{\bf{n}}_k}^{-1}{\bf{H}}_k{\bf{F}}_k \succ {\bf{F}}^{\rm{H}}_{k,{\rm{PO}}}{\bf{H}}_{k}^{\rm{H}}
{\bf{R}}_{{\bf{n}}_k}^{-1}{\bf{H}}_k{\bf{F}}_{k,{\rm{PO}}}$.
The Pareto optimal solution set of ${\bf{F}}_{k}$ of \textbf{Prob. 1} consists of the optimal ${\bf{F}}_{k}$ of the following optimization problem by traversing all possible ${\bf{F}}_{k,{\rm{PO}}}$
\begin{align}
& {\textbf{Prob. 2:}}\nonumber \\
  &  \max_{\alpha,{\bf{F}}_k} \ \ \ \ \ \ \ \ \ \ \ \  \alpha \nonumber \\
& \ \ \ {\rm{s.t.}} \ \ \ \  {\bf{F}}^{\rm{H}}_k{\bf{H}}_{k}^{\rm{H}}
{\bf{R}}_{{\bf{n}}_k}^{-1}{\bf{H}}_k{\bf{F}}_k ={\alpha}{\bf{F}}^{\rm{H}}_{k,{\rm{PO}}}{\bf{H}}_{k}^{\rm{H}}
{\bf{R}}_{{\bf{n}}_k}^{-1}{\bf{H}}_k{\bf{F}}_{k,{\rm{PO}}}\nonumber \\
& \ \ \ \ \ \ \ \ \ \ \ {\rm{Tr}}({\bf{F}}_k{\bf{F}}_k^{\rm{H}})\le P_k \nonumber \\
&¡¡\ \ \ \ \ \ \ \ \ \ \ {\bf{F}}_k{\bf{F}}_{k}^{\rm{H}} \preceq \tau_{k,{\rm{max}}}{\bf{I}}.
\end{align}When computing optimal ${\bf{F}}_k$ and $\alpha$, ${\bf{F}}_{k,{\rm{PO}}}$ is a given matrix instead of an unknown matrix.
 It is proved in Appendix~\ref{Optimal_Structure} regardless of the specific values of ${\bf{F}}_{k,{\rm{opt}}}$ the optimal solution of ${\bf{F}}_k$ of \textbf{Prob. 2} always satisfies the following structure
\begin{align}
\label{Optimal_Structure_A}
{\bf{F}}_{k,{\rm{opt}}}={\bf{V}}_{{\boldsymbol {\mathcal H}}_k}{\boldsymbol \Lambda}_{{\bf{F}}_k}{\bf{U}}_{{\rm{Arb}},k}^{\rm{H}} \ \ {\rm with}  \ \  [{\boldsymbol \Lambda}_{{\bf{F}}_k}]_{i,i} \le \tau_{\max}
 \end{align}where the unitary ${\bf{V}}_{{\boldsymbol{\mathcal{H}}}_k}$ is  defined based SVD ${\bf{R}}_{{\bf{n}}_k}^{-1/2}{\bf{H}}_k={\bf{U}}_{{\boldsymbol{\mathcal{H}}}_k}
{\boldsymbol \Lambda}_{{\boldsymbol{\mathcal{H}}}_k}{\bf{V}}_{{\boldsymbol{\mathcal{H}}}_k}^{\rm{H}}$ with ${\boldsymbol \Lambda}_{{\boldsymbol{\mathcal{H}}}_k} \searrow $. Additionally. the first $N$ diagonal elements of the diagonal matrix ${\boldsymbol \Lambda}_{{\bf{F}}_k}$ are still unknown variables and the other diagonal elements are all zeros. As the Pareto optimal solution set of ${\bf{F}}_k$ of \textbf{Prob. 1} can be achieved by the resulting optimal solution of {\textbf{Prob. 2}} via changing ${\bf{F}}_{k,{\rm{opt}}}$, it can be directly concluded that any Pareto optimal solution of \textbf{Prob. 1} satisfies the optimal structure given by (\ref{Optimal_Structure_A}). Based on the optimal structure the remaining problem becomes how to compute ${\boldsymbol \Lambda}_{{\bf{F}}_k}$, and it will be discussed in the following section.

\subsection{Cave Water-filling Solutions}

At the beginning of this section, we want to highlight that the solutions of ${\boldsymbol \Lambda}_{{\bf{F}}_k}$ in (\ref{Optimal_Structure_A}) are determined by the specific formulas of  objective function. Different objective functions usually have different optimal ${\boldsymbol \Lambda}_{{\bf{F}}_k}$. However, the derivation logics for different objective functions are exactly the same, which all exploit the famous Karush-Kuhn-Tucker (KKT) conditions and the final solutions are variants of classic water-filling solutions, which are termed as cave water-filling solutions \cite{FFGao08}. In the following two most representative objectives are investigated, i.e., A-Schur-Convex and M-Schur-Convex objective functions. For these objective functions, the optimal solutions are independent of the specific formulations of the objective functions \cite{Palomar03,Palomar2007,XingTSP2014,JSAC_Xing2012}. In terms of BER, these objective functions usually enjoy much better performance over their corresponding Schur-concave counterparts, respectively \cite[P.385]{Palomar2007}.

\noindent \textbf{M-Schur-Convex:}

For M-Schur-Convex objective functions, based on the optimal ${\bf{Q}}_k$'s in (\ref{Q_0_opt}) and (\ref{Q_k_opt}) the objective function in \textbf{Obj. 5} is equivalent to minimizing $\prod_{i=1}^N\lambda_i({\boldsymbol { \Phi}}_{\rm{LMMSE}}(\{{\bf{Q}}_k\},\{{\bf{F}_k}\}))$ \cite{JSAC_Xing2012}. Based on the optimal structure given by (\ref{Optimal_Structure_A}) and SVD ${\bf{R}}_{{\bf{n}}_k}^{-1/2}{\bf{H}}_k={\bf{U}}_{{\boldsymbol{\mathcal{H}}}_k}
{\boldsymbol \Lambda}_{{\boldsymbol{\mathcal{H}}}_k}{\bf{V}}_{{\boldsymbol{\mathcal{H}}}_k}^{\rm{H}}$ with ${\boldsymbol \Lambda}_{{\boldsymbol{\mathcal{H}}}_k} \searrow $, defining
\begin{align}
[{\boldsymbol \Lambda}_{{\bf{F}}_k}]_{i,i}=f_{k,i}, \ \ [{\boldsymbol \Lambda}_{{\boldsymbol{\mathcal{H}}}_k}]_{i,i}=h_{k,i}
\end{align}
and substituting (\ref{Optimal_Structure_A}) into $\prod_{i=1}^N\lambda_i({\boldsymbol { \Phi}}_{\rm{LMMSE}}(\{{\bf{Q}}_k\},\{{\bf{F}_k}\}))$  the transceiver designs with M-Schur-Convex objective functions is equivalent to
\begin{align}
\label{opt:capacity}
& \min_{f_{k,i}^2} \ \ \ \sum_{i=1}^{N}{\rm{log}} \left(1- \frac{\prod_{k=1}^Kf_{k,i}^2h_{k,i}^2}{\prod_{k=1}^K(f_{k,i}^2h_{k,i}^2+1)}\right) \nonumber \\
& \ {\rm{s.t.}} \ \ \ \ \sum_{i=1}^N f_{k,i}^2 \le P_k \nonumber \\
& \ \ \ \ \ \ \ \ \ \ f_{k,i}^2 \le \tau_{k,{\max}}.
\end{align}Generally, the considered optimization problem (\ref{opt:capacity}) is nonconvex and thus it is difficult to derive the closed-form optimal solutions. Following the logic in \cite{Yu04}, an iterative algorithm is further exploited to solve the unknown variables. Defining the following auxiliary variable
\begin{align}
\label{a_k_i}
 a_{k,i}=\prod_{l\not=k}\frac{f_{l,i}^2h_{l,i}^2}{f_{l,i}^2h_{l,i}^2+1}
\end{align}the Lagrangian of (\ref{opt:capacity}) is of the following form
\begin{align}
&\mathcal{L}(\{f_{k,i}^2\},\mu_k,\{\gamma_{k,i}\},\{l_{k,i}\})\nonumber \\
=&\sum_{i=1}^{N}{\rm{log}} \left(1- a_{k,i}\frac{f_{k,i}^2h_{k,i}^2}{f_{k,i}^2h_{k,i}^2+1}\right) +\mu_k(\sum_{i=1}^N f_{k,i}^2 - P_k)\nonumber \\
&+\sum_{i=1}^N\{\gamma_{k,i}(f_{k,i}^2 -\tau_{k,{\max}})\}
-\sum_{i=1}^N l_{k,i} f_{k,i}^2,
\end{align} where $f_{k,i}^2$'s are taken as the variables and then there are hidden constraints that $f_{k,i}^2\ge 0$, which lead to the final term. Based on the KKT conditions
we have the follow equation
\begin{align}
&\mu_k+\gamma_{k,i}-l_{k,i}\nonumber \\
=&\underbrace{\frac{a_{k,i}h_{k,i}^2}{(1- a_{k,i})(f_{k,i}^2h_{k,i}^2+1)^2+{a_{k,i}}{(f_{k,i}^2h_{k,i}^2+1)}}}
_{\triangleq \mathcal{\hat H}_{k,i}(f_{k,i}^2)}
\end{align}based on which $f_{k,i}^2$ can be solved to be \begin{align}\label{f_M_Schur_Convex}
&f_{k,i}^2=\left\{ {\begin{array}{*{20}{c}}
 \ \ {\mathcal{F}_{k,i} \ \ \ \ \mu_k \ge \mathcal{\hat H}_{k,i}(\tau_{k,\max})}\\
{ \tau_{k,{\max}} \ \ \ \  \mu_k < \mathcal{\hat H}_{k,i}(\tau_{k,\max}) }
\end{array}} \right. \nonumber \\
&{\rm{with}}\nonumber \\
 &{\mathcal{F}}_{k,i}=\frac{1}{h_{k,i}^2}\left( \frac{-a_{k,i}+
\sqrt{a_{k,i}^2+4(1-a_{k,i})a_{k,i}h_{k,i}^2/\mu_k}}{2(1-a_{k,i})}-1 \right)^{+}
\end{align}where the Lagrange multiplier $\mu_k$ makes sure $\sum_i f_{k,i}^2=P_k$.  The conditions $\mu_k \ge \mathcal{\hat H}_{k,i}(\tau_{k,\max})$ and $\mu_k < \mathcal{\hat H}_{k,i}(\tau_{k,\max})$ mean that the power invested on each eigenchannel cannot exceed $\tau_{k,{\max}}$.

\noindent \textbf{A-Schur-Convex}

On the other hand, based on the optimal ${\bf{Q}}_k$'s in (\ref{Q_0_opt}) and (\ref{Q_k_opt}) the objective function in \textbf{Obj. 3} is equivalent to minimizing $\sum_{i=1}^N\lambda_i({\boldsymbol { \Phi}}_{\rm{LMMSE}}(\{{\bf{Q}}_k\},\{{\bf{F}_k}\}))$ \cite{XingTSP2013}. Substituting the optimal structure in (\ref{Optimal_Structure_A}) into $\sum_{i=1}^N\lambda_i({\boldsymbol { \Phi}}_{\rm{LMMSE}}(\{{\bf{Q}}_k\},\{{\bf{F}_k}\}))$ the optimization problem with A-Schur-Convex objective functions is equivalent to the following problem
\begin{align}
\label{opt_scalar_MMSE}
&\min_{f_{k,i}^2} \ \sum_{i=1}^N\left(1-\frac{\prod_{k=1}^K f_{k,i}^2h_{k,i}^2}
{\prod_{k=1}^K (f_{k,i}^2h_{k,i}^2+1)}\right) \nonumber \\
 &\ {\rm{s.t.}} \ \   \sum_{i=1}^N f_{k,i}^2 \le P_k \nonumber \\
 & \ \ \ \ \ \ \  f_{k,i}^2 \le \tau_{k,{\max}}.
\end{align}Similarly, an iterative algorithm is exploited to solve $f_{k,i}^2$'s. At each iteration, the Lagrangian of (\ref{opt_scalar_MMSE}) equals
\begin{align}
&\mathcal{L}(\{f_{k,i}^2\},\mu_k,\{\gamma_{k,i}\},\{l_{k,i}\})\nonumber \\=&\sum_{i=1}^N\left(1-\frac{\prod_{k=1}^K f_{k,i}^2h_{k,i}^2}
{\prod_{k=1}^K (f_{k,i}^2h_{k,i}^2+1)}\right)+\mu_k(\sum_{i=1}^N f_{k,i}^2 - P_k)\nonumber \\
&+\sum_i\{\gamma_{k,i}(f_{k,i}^2 -\tau_{k,{\max}})\}-\sum_i l_{k,i} f_{k,i}^2.
\end{align} Based on its KKT conditions, we will obtain the following equation with respect to $f_{k,i}^2$
\begin{align}
\mu_k+\gamma_{k,i}-l_{k,i}=\frac{a_{k,i}h_{k,i}^2}
{(h_{k,i}^2f_{k,i}^2+1)^2}\triangleq {\mathcal {\tilde H}}_{k,i}(f_{k,i}^2)
\end{align} based on which $f_{k,i}^2$ can be solved to be
\begin{align}\label{f_A_Schur_Convex}
f_{k,i}^2=\left\{ {\begin{array}{*{20}{c}}
{\left(\sqrt{\frac{a_{k,i}}{\mu_kh_{k,i}^2 }}
-\frac{1}{h_{k,i}^2}\right)^{+} \ \ \ \ \ \mu_k \ge {\mathcal{\tilde H}}_{k,i}(\tau_{k,\max})}\\
{\ \ \ \ \ \ \ \   \tau_{k,{\max}} \ \ \ \ \ \ \ \ \ \ \ \ \ \ \ \mu_k < {\mathcal{\tilde H}}_{k,i}(\tau_{k,\max})}
\end{array}} \right.
\end{align}where the Lagrange multiplier $\mu_k$ makes sure $\sum_i f_{k,i}^2=P_k$. Additionally, the conditions $\mu_k \ge \mathcal{\tilde H}_{k,i}(\tau_{k,\max})$ and $\mu_k < \mathcal{\tilde H}_{k,i}(\tau_{k,\max})$ also mean that the power invested on each eigenchannel cannot be larger than $\tau_{k,{\max}}$.

Due to the fact that there are ceiling constraints on each eigenchannel, the previous results can be named as cave water-filling solutions \cite{FFGao08}. In order to show clearly how to compute the cave water-filling solutions, a detailed diagram for the implementation of the previous cave water-filling solutions is given in \textbf{Algorithm 1}. It is worth noting that \textbf{Algorithm 1} is not restricted to the specific formulas of cave water-filling solutions.
\begin{algorithm}
\caption{Cave Water-Filling Algorithm}
\begin{algorithmic}[1]

\STATE $\mathcal{C}=\{\text{Indexes}$ $\text{of}$ $\text{all}$ $\text{the}$ $\text{eigenchannels}$\}

\STATE $\text{Initialize}:$ $\text{Take} \ \text{waterfilling} \ \text{of} \ P_k \  \text{over} \ \mathcal{C} \ \text{neglecting} \ \ \tau_{k,{\max}}.$

\WHILE{$\text{length}( \text{find}(\text{PowerAllo}>\tau_{k,{\max}} )) > 0$}
\STATE $\{\text{Indexmax}\}=\text{Indexes}$ $\text{of}$ $\text{find}$($\text{PowerAllo} >\tau_{k,{\max}})$;
\STATE $\{\text{Indexnormal}\}=\mathcal{C}/\{\text{posIndex}\}( \text{exclusion} \ \text{operation} )$;
\STATE $\text{Set}$ $\text{PowerAllo}(\text{Indexmax})=\tau_{k,{\max}}$;
\STATE $\text{Set}$ $L=\text{Cardinality} \ \text{of} \ \{\text{Indexmax}\}$;

\STATE $\text{Take} \ \text{waterfilling} \ \text{of}$ $P_k-L\tau_{k,{\max}}$ $\text{over}$  $\{\text{Indexnormal}\}$ $\text{without} \ \text{considering} \ \tau_{k,{\max}s}.$

\ENDWHILE

\RETURN $\text{PowerAllo}$

\end{algorithmic}
\end{algorithm}
\noindent \textbf{Distributed Implementation}

Based on the optimal structure of ${\bf{F}}_k$ in (\ref{Optimal_Structure_A}) and optimal ${\bf{Q}}_k$ in (\ref{Q_k_opt}), exploiting the definition of ${\bf{F}}_k$ in (\ref{F_definition}) the optimal forwarding matrix at the $k^{\rm{th}}$ node satisfies ${\bf{P}}_{k,{\rm{opt}}}={\bf{V}}_{{\boldsymbol {\mathcal H}}_k}{\boldsymbol \Lambda}_{{\bf{P}}_k}{\bf{U}}_{{\boldsymbol {\mathcal H}}_{k-1}}^{\rm{H}}$ where ${\boldsymbol \Lambda}_{{\bf{P}}_k}$ is a diagonal matrix. The matrices ${\bf{V}}_{{\boldsymbol {\mathcal H}}_k}$ and ${\bf{U}}_{{\boldsymbol {\mathcal H}}_{k-1}}^{\rm{H}}$ are determined by the CSI of the neighboring preceding and succeeding links. It is easy to gain such kind of CSI.  The computation of the diagonal elements of ${\boldsymbol \Lambda}_{{\bf{P}}_k}$ needs to share the information of ${\boldsymbol \Lambda}_{{\bf{P}}_k}$'s. It should be highlighted that based on the previous discussions, the computation of  ${\boldsymbol \Lambda}_{{\bf{P}}_k}$ only needs the auxiliary variables $\{{{a}}_{k,i}\}_{i=1}^{N}$ given in  (\ref{a_k_i}). It means that only a $N$-dimensional vector needs to be shared between the neighbouring nodes.

\noindent {\textbf{Remark:}} When channel estimation errors are taken into account, we will have ${\bf{H}}_k={\bf{\hat H}}_k+\Delta{\bf{H}}_k$ where ${\bf{\hat H}}_k$ is the channel estimation in the $k^{\rm{th}}$ hop and $\Delta{\bf{H}}_k$ is the corresponding channel estimation error. If the elements of $\Delta{\bf{H}}_k$ are i.i.d. random variables with mean zero and variance $\sigma_{e_k}^2$, our proposed solutions for the transceiver designs under joint power constraints can be directly to extend to a robust design by simply replacing ${\bf{H}}_k$ by ${\bf{\hat H}}_k$ and $\sigma_{n_k}^2$ by $\sigma_{n_k}^2+P_{k}\sigma_{e_k}^2$.

\section{Discussions on The Proposed Solutions}
In this paper, we have investigated in depth two special cases of (\ref{Optimziation_Original}), i.e., the transceiver designs with pure shaping constraints and with joint power constraints. Both the two special cases have clear physical meanings \cite{Palomar2004,Dai2012}. Based on the previous derivations and conclusions, it is interesting that in the case with pure shaping constraints, the proposed solution of ${\bf{F}}_k$ is independent of the objective functions. In other words, in this case for Objs. 1-6 in Section~\ref{Problem_Formulation} ${\bf{F}}_k$ have the same solution (\ref{F_k_pure}). On the other hand, in the case with joint power constraints, the optimal solutions of ${\bf{F}}_k$ for Objs. 1-6 in Section~\ref{Problem_Formulation} have the same structure given by (\ref{Optimal_Structure_A}). However, there is still an unknown diagonal matrix ${\boldsymbol \Lambda}_{{\bf{F}}_k}$ that still needs to optimized. The optimal solution for this diagonal matrix is problem dependent, but the logics to derive optimal solutions are the same for Objs. 1-6, which are based KKT conditions.

It should be highlighted that although six kinds of objective functions are considered, the number of the specific objective functions for Objs. 1-6 seems  infinite. For example, there are many functions that all are A-Schur-Concave. The stories are similar for A-Schur-Convex, M-Schur-Concave and M-Schur-Convex objective functions. It is worth noting that  for Objs. 1-6 in Section~\ref{Problem_Formulation} the logic for deriving the optimal ${\boldsymbol \Lambda}_{{\bf{F}}_k}$ is the same. Therefore, in our work only two most representative objectives are chosen, which are A-Schur-Convex and M-Schur-Convex, because when the objective functions are A-Schur-Convex or M-Schur-Convex the optimal ${\boldsymbol \Lambda}_{{\bf{F}}_k}$ is independent of the specific formulas of the objective functions.

The logic in Subsection C of  Section~\ref{Joint_Power_Constraint} can be directly applied to the other objective functions.
Meanwhile, the solutions in Subsection C of  Section~\ref{Joint_Power_Constraint} also have a broader range of applications.
For example, for the capacity maximization problem the objective function can be considered to be M-Schur-Concave and in this case expect a different ${\bf{Q}}_1$ the optimal solution of $f_{k,i}^2$ is exactly the same as the one for M-Schur-Convex (\ref{f_M_Schur_Convex}). The objective function sum MSE minimization can be taken as an A-Schur-Concave objective function and similarly except a different ${\bf{Q}}_1$ the optimal solution of $f_{k,i}^2$ is exactly the same as the one for A-Schur-Convex (\ref{f_A_Schur_Convex}). In our work, the iterative algorithms named iterative cave water-filling are adopted to solve $f_{k,i}^2$. Moreover, for the iterative algorithms, at each iteration the optimization is convex and for the optimal solutions the objective functions decrease monotonically. The convexity guarantees the convergence of the proposed iterative algorithms.

Our works can degrade to  several existing works.

\begin{itemize}

\item  When the covariance shaping constraints is inactive and linear transceiver is adopted, our work will become linear transceiver designs for multi-hop AF MIMO relaying systems.

\item When the covariance shaping constraints is inactive and nonlinear transceiver is adopted, our work will become nonlinear transceiver designs with THP or DFE for multi-hop AF MIMO relaying systems.

\item When the covariance shaping constraints is inactive and the number of hops is set to be 2, our work will become the traditional linear or nonlinear transceiver designs under sum power constraint for dual-hop AF MIMO relaying systems.

\item In the dual-hop case, if the noise variance at relay tends to be 0 and covariance shaping constraints is removed, our work will become to be the traditional transceiver designs under sum power constraint for point-to-point MIMO systems.

\item In the dual-hop case, if the noise variance at relay tends to be 0, the transceiver designs with pure shaping constraints and joint power constraints will become to be their counterparts for point-point MIMO systems, respectively.

\end{itemize}

Finally, for the general optimization problem of (\ref{Optimziation_Original}), i.e., all constraints are active and ${\bf{R}}_{{\bf{s}}_k}$ is not identity,  solving its  subproblem (\ref{subproblem_matrix}) is very challenging. To the best of the authors' knowledge even for the simple point-to-point MIMO systems, the transceiver design problem is still largely open. By the way, comparing the optimal structures of ${\bf{F}}_k$ for the two cases with shaping constraints and joint power constraints, the two optimal structures are significantly different. In specific, the former one is channel independent, but the latter one is channel dependent. It will be very challenging to unify them in a single formulation, but it is a very good direction for future research .

\section{Simulation Results and Discussions}
\label{simulation}
In this section, the performance of the transceiver designs under covariance shaping constraints will be evaluated by simulations. Without loss of generality, a three-hop AF MIMO relaying networks is simulated, in which there is one source, two relays and one destination. In addition, it is also assumed that all nodes are equipped with 4 antennas. The entries of the channel matrix in each hop are i.i.d. circularly symmetric complex Gaussian distributed with zero mean and unit variance. The source node aims to transmit four independent data streams to the destination.  In the following figures, each point is an average of 2000 independent trials. The signal-to-noise ratio (SNR) in each hop is defined as ${\rm{SNR}}_k=P_{k}/\sigma_{n_k}^2$ and for simplicity in our simulations it is set that $P_1=P_2=P_3=4$ and ${\rm{SNR}}_1={\rm{SNR}}_2={\rm{SNR}}_3={\rm{SNR}}$ (different SNRs are realized by adjusting noise variances).

Firstly, the performance of the transceiver designs with pure shaping constraints is assessed by the simulations. In this case, it is a natural problem how to choose ${\bf{R}}_{{\bf{s}}_k}$'s in shaping constraints.
In the existing work \cite{Palomar2004}, the covariance shaping matrices are assumed to be known a priori. Regarding their explicit formulas, only the simplest diagonal structure is given but without any theoretical analysis and simulation to support it. In this simulation we want to model per-antenna power constraints by constructing ${\bf{R}}_{{\bf{s}}_k}$'s, and the diagonal elements of ${\bf{R}}_{{\bf{s}}_k}$ must be equal to or smaller than some prescribed threshold values. In our setting, threshold values are set to be $[0.4 \ 0.8 \ 1.2 \ 1.6]$ without loss of generality.
In the following, two detailed schemes are proposed to construct ${\bf{R}}_{{\bf{s}}_k}$'s.

The design of ${\bf{R}}_{{\bf{s}}_k}$ should be much better if it is independent of specific channel realizations.
As in nature ${\bf{R}}_{{\bf{s}}_k}$ can be understood as a correlation matrix, the first one is to exploit the well-known exponential structure as follows and denoted
\begin{figure}[!ht]
\centering
\includegraphics[width=.34\textwidth]{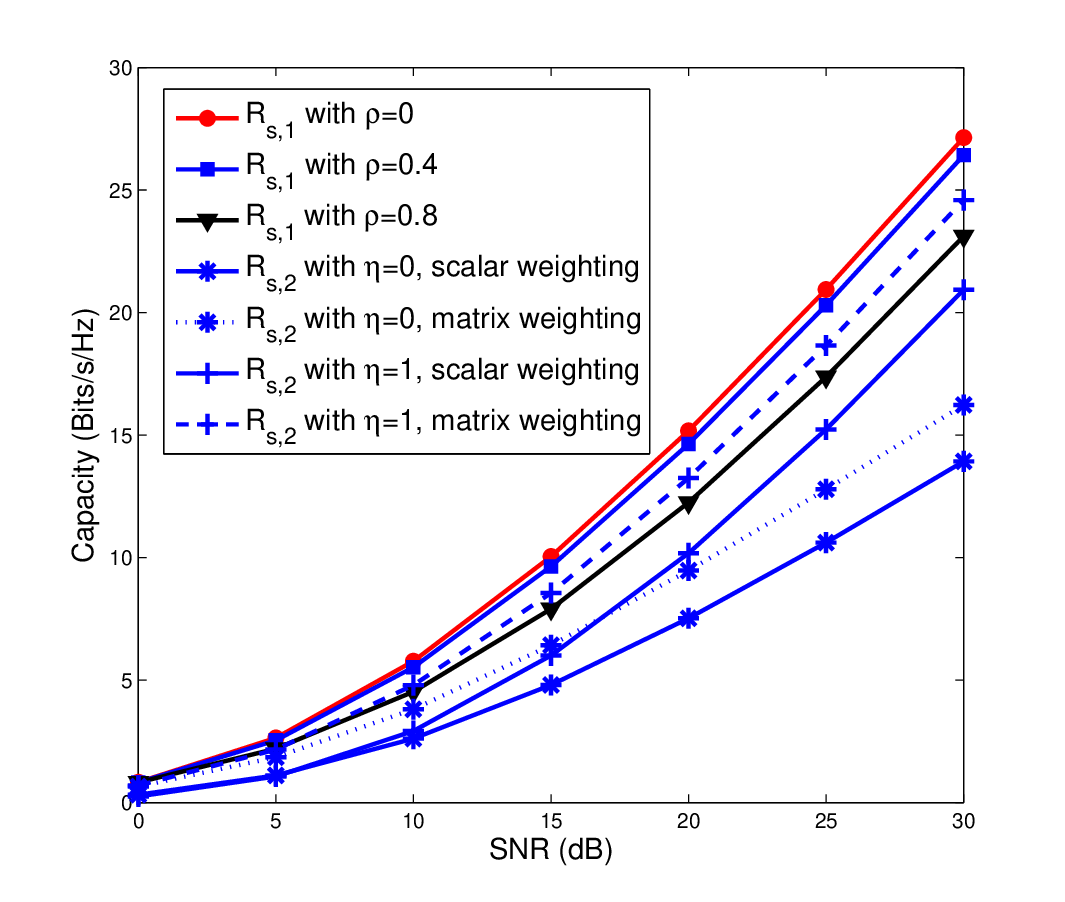}
\caption{The channel capacities under various shaping constraints}\label{fig:5}
\end{figure} by ${\bf{R}}_{{\bf{s}},1}$
\begin{align}
&{\bf{R}}_{{\bf{s}},1}={\rm{diag}}\{\{\sqrt{p_j}\}_{j=1}^4\}\left[ {\begin{array}{*{20}{c}}
1&{\rho }&{\rho^2}&{\rho^3 }\\
{\rho }&1&{\rho }&{\rho^2}\\
{\rho^2}&{\rho}&1&{\rho }\\
{\rho^3}&{\rho^2}&{\rho }&1
\end{array}} \right]\nonumber \\
& \ \ \ \ \ \ \ \ \ \ \ \times{\rm{diag}}\{\{\sqrt{p_j}\}_{j=1}^4\}
\end{align}where $0\le \rho <1 $ is the exponential factor. The term ${\rm{diag}}\{\{\sqrt{p_j}\}_{j=1}^4\}$ aims at making sure the diagonal elements equal to the prescribed thresholds and meanwhile guaranteeing ${\bf{R}}_{{\bf{s}}_k}$ is a Hermitian matrix via simple linear operations.

To make a comparison with the exponential structure another choice is to produce ${\bf{R}}_{{\bf{s}}_k}$ based on the following special structure ${\bf{R}}_{{\bf{s}},2}$ whose eigenvectors are the same as those of the corresponding channel matrix
\begin{align}
{\bf{R}}_{{\bf{s}},2}={\bf{V}}_{{\boldsymbol{\mathcal{H}}}_k}{\boldsymbol {\Lambda}}_{k}{\bf{V}}_{{\boldsymbol {\mathcal{H}}}_k}^{\rm{H}},
\end{align}where the diagonal entries of the diagonal matrix ${\boldsymbol {\Lambda}}_{k}$ are adjustable. The produce is in nature to solve a linear equation array, that adjusting ${\boldsymbol {\Lambda}}_{k}$ makes the diagonal elements equivalent to the threshold values. Unfortunately, there is a problem. From physical meaning, ${\bf{R}}_{{\bf{s}}_k}$ must be positive semi-definite, however this fact cannot always hold in the computation of ${\boldsymbol {\Lambda}}_{k}$. In other words, after computation some diagonal entries of ${\boldsymbol {\Lambda}}_{k}$  may be negative. To overcome this problem, here two methods are adopted. One is to simply set the negative diagonal entries to be zeros. The other is to set them to a positive value $\eta$ instead of zero, which is designed empirically. Followup, these two operations may violate the constraints on the diagonal elements of ${\bf{R}}_{{\bf{s}}_k}$ as they make ${\bf{R}}_{{\bf{s}}_k}$ more positive. As a result, two weighting operations are exploited further to make the constraints satisfied, which are named matrix weighting and scalar weighting, respectively:
\begin{align}
&{\textbf{Matrix \ Weighting :}} \ \ \nonumber \\ &{\bf{R}}_{{\bf{s}},2}={\rm{diag}}\{\{\sqrt{\beta_j}\}_{j=1}^4\} {\bf{V}}_{{\boldsymbol{\mathcal{H}}}_k}{\boldsymbol {\Lambda}}_{k}{\bf{V}}_{{\boldsymbol {\mathcal{H}}}_k}^{\rm{H}}{\rm{diag}}\{\{\sqrt{\beta_j}\}_{j=1}^4\} \\
&{\textbf{Scalar \ Weighting:}} \ \ \nonumber \\
&{\bf{R}}_{{\bf{s}},2}=\beta {\bf{V}}_{{\boldsymbol{\mathcal{H}}}_k}{\boldsymbol {\Lambda}}_{k}{\bf{V}}_{{\boldsymbol {\mathcal{H}}}_k}^{\rm{H}},
\end{align}where $\beta$ is the maximum scalar that makes the constraints on the diagonal elements satisfied.

\begin{figure}[!ht]
\centering
\includegraphics[width=.34\textwidth]{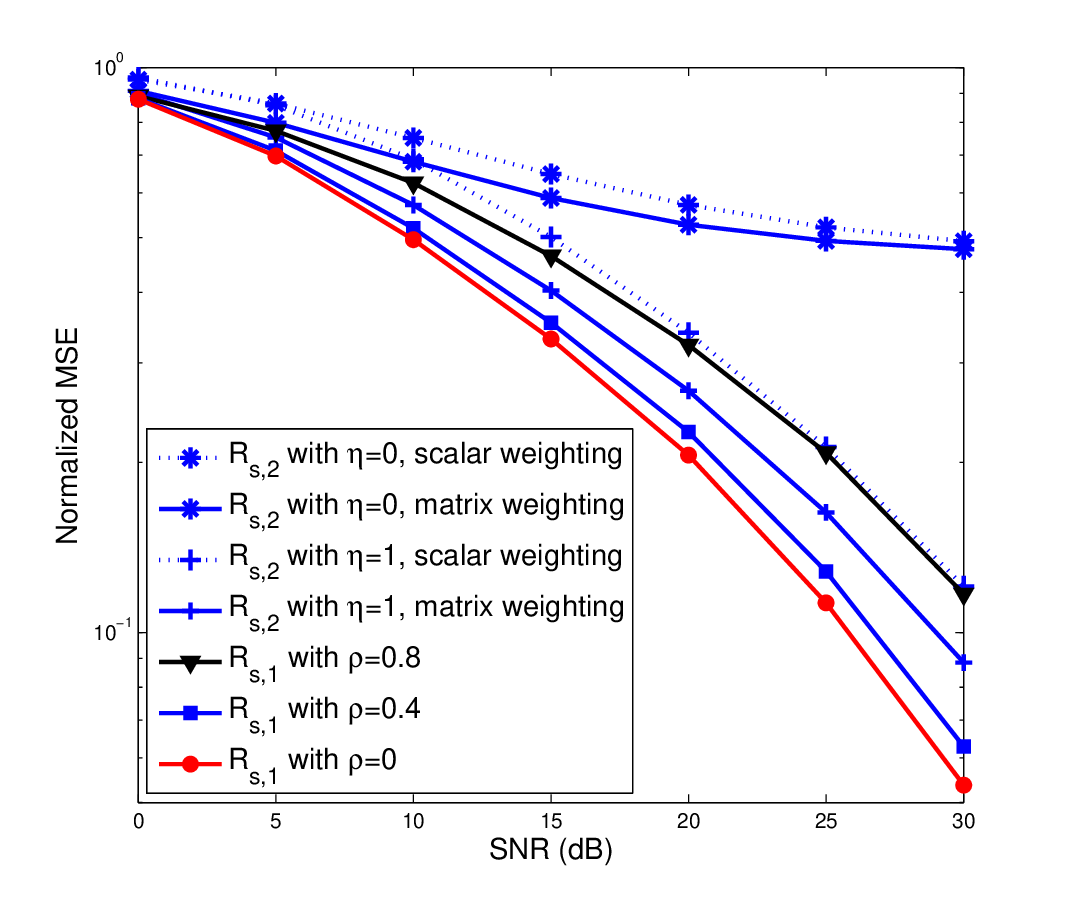}
\caption{The normalized MSEs under various shaping constraints}\label{fig:6}
\end{figure}

Based on the previous two schemes for ${\bf{R}}_{{\bf{s}}_k}$, the performance of the transceiver designs with pure shaping constraints investigated. In this case, the optimal transceiver structure is given by (\ref{F_k_pure}). The capacities of the different designs are plotted in Fig.~\ref{fig:5}.
In Fig.~\ref{fig:5} we can see an interesting result that the best performance is achieved when choosing ${\bf{R}}_{{\bf{s}},1}$ with $\rho=0$.
In other words, as shown by the simulation results the simplest diagonal structure is the best.
As $\rho$ increases, the capacity decreases. The performance for ${\bf{R}}_{{\bf{s}},2}$ with $\eta=0$ is very poor. This is because it closes some eigenchannels for transceiver designs. On the other hand, in our simulation we changes the values of $\eta$ and discover that when $\eta=1$ much better performance can be obtained. Furthermore, referring to ${\bf{R}}_{{\bf{s}},2}$ matrix weighting operation always outputs scalar weighting operation although scalar weighting operation can keep the eigenvectors of ${\bf{R}}_{{\bf{s}},2}$ unchanged. Based on this result, we may argue that whether eigenvectors of ${\bf{R}}_{{\bf{s}},2}$ match the channel or not is not important.  Similar conclusions can also be achieved for sum MSE minimization as depicted in Fig.~\ref{fig:6} with Quadrature Phase Shift Keying (QPSK) being chosen as the modulation constellation.

\begin{figure}[!ht]
\centering
\includegraphics[width=.34\textwidth]{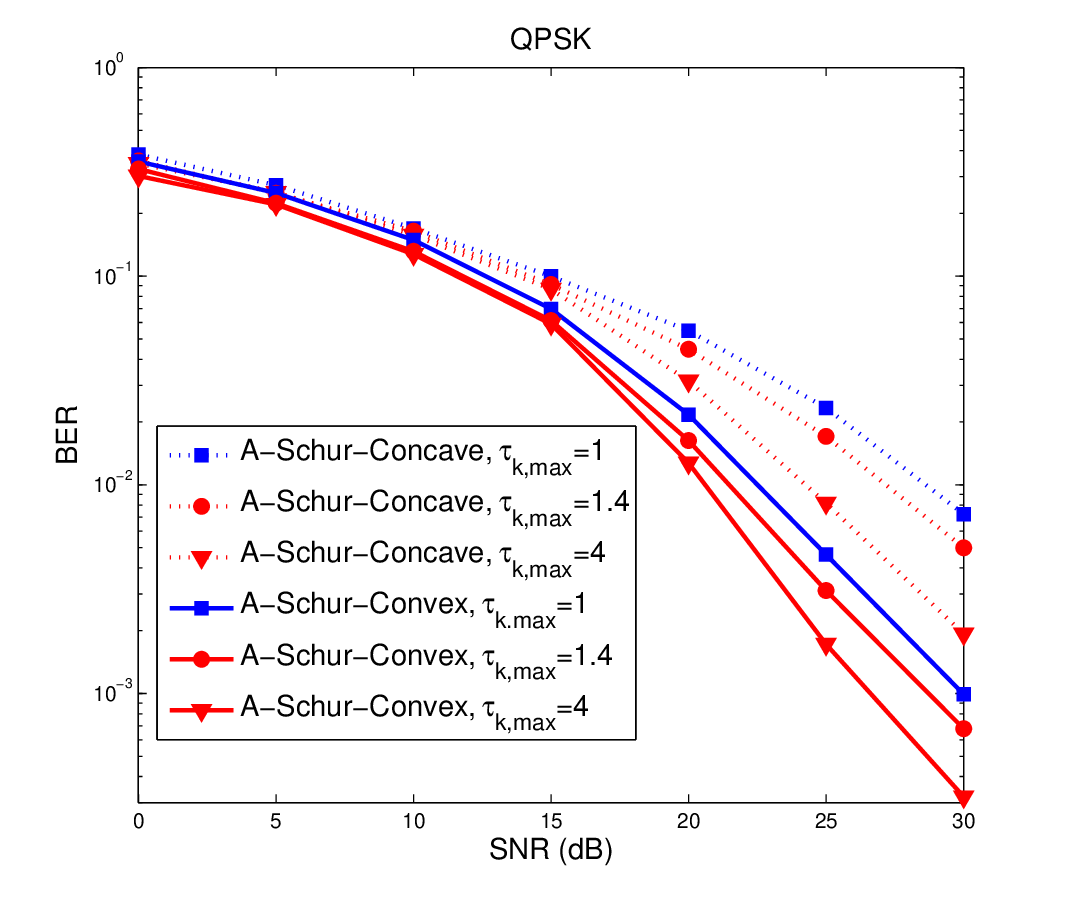}
\caption{The BERs of linear transceiver designs with A-Schur-Convex and A-Schur-Concave objectives for different $\tau_{k,\max}$.}\label{fig:7}
\end{figure}

\begin{figure}[!ht]
\centering
\includegraphics[width=.34\textwidth]{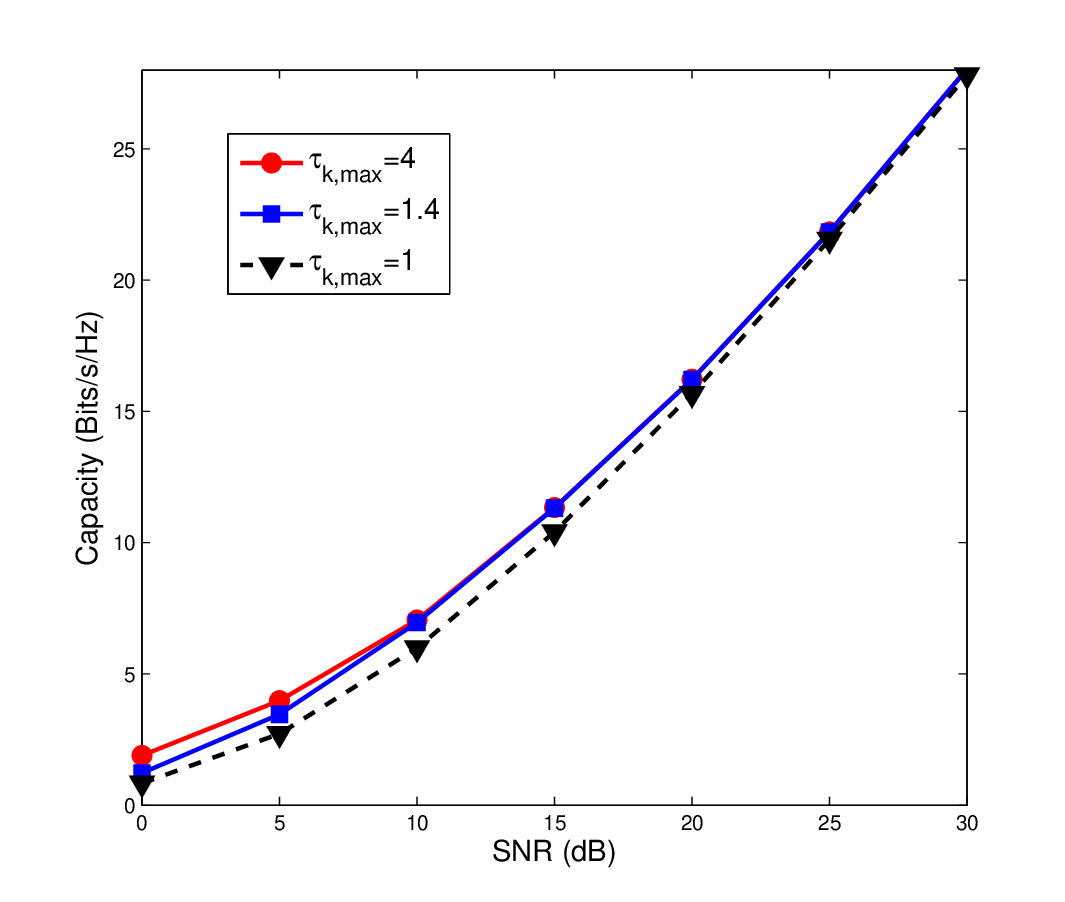}
\caption{The channel capacities under different peak power constraints.}\label{fig:8}
\end{figure}

When joint power constraints are taken into account, the effects of peak power constraints are first evaluated by simulation results. The BERs of the linear transceiver designs with different objectives are shown in Fig.~\ref{fig:7} when QPSK is used as the modulation constellation. Both A-Schur-Convex and A-Schur-Concave objective functions are simulated here, which correspond MAX-MSE minimization and sum-MSE minimization, respectively \cite[P.385]{Palomar2007}. It is obvious that the tighter the joint power constraint is the worse performance the designs have. Additionally, the linear transceiver with A-Schur-Convex objective function always has a much better performance in terms of BER than its counterpart with A-Schur-Concave objective function. While for capacity maximization as shown in Fig.~\ref{fig:8},  the story is a little bit different as in high SNR regime, the designs with different peak power constraints have almost the same performance. It is because for capacity maximization, if the sum power constraint is active, at high SNR the optimal solution is to allocate the power uniformly, which will be independent of the peak power constraints.

\begin{figure}[!ht]
\centering
\includegraphics[width=.34\textwidth]{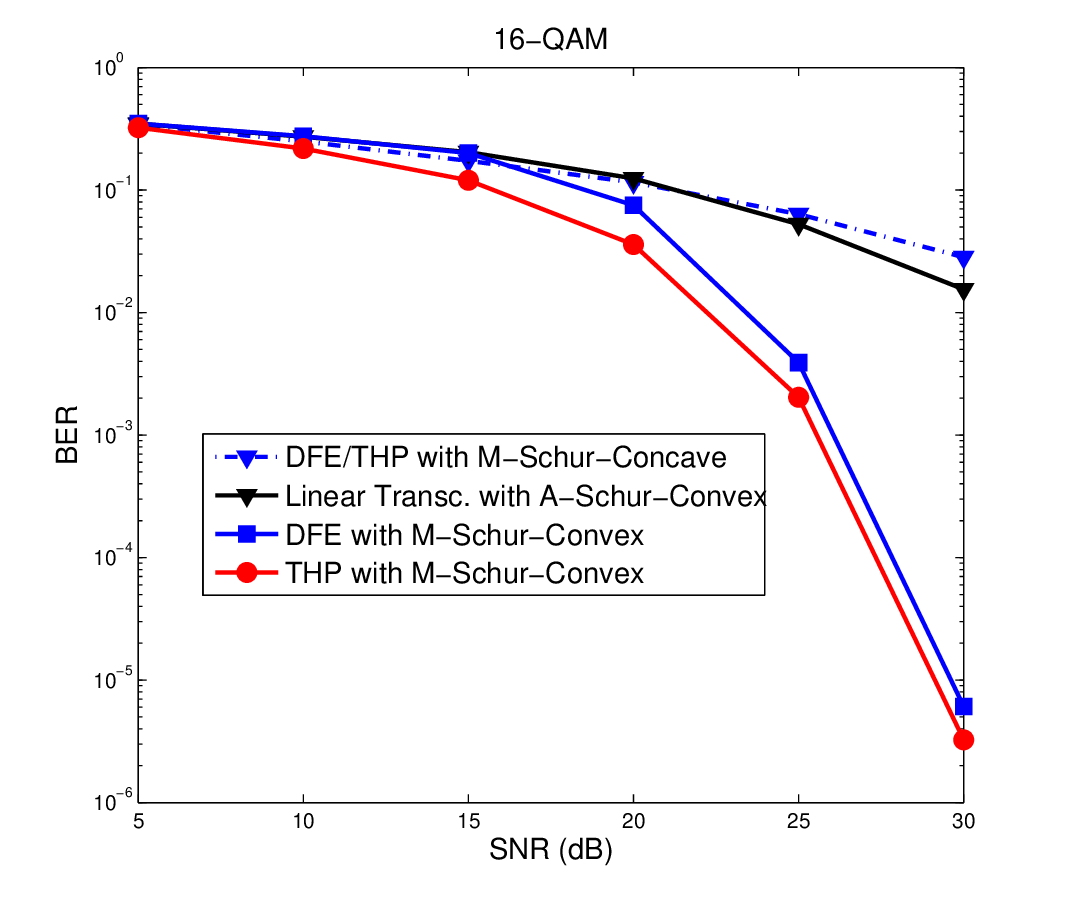}
\caption{The BERs of the nonlinear and linear transceiver designs under joint power constraints with $\tau_{k,\max}$=1.4.}\label{fig:9}
\end{figure}

Furthermore, when THP or DFE is adopted to improve BER at the cost of high complexity, high order modulation is preferred.
Under the joint constraints with $\tau_{k,{\max}}=1.4$ the BERs of various transceiver designs are further compared in Fig.~\ref{fig:9} when 16 Quadrature Amplitude Modulation (16-QAM) constellation is used. For the curves in Fig.~\ref{fig:9}, Gray code is  adopted to further improve the BER performance. Note that when the objective function is M-Schur-Concave, the THP and DFE structures will both reduce to linear capacity maximization transceiver \cite{JSAC_Xing2012}. For A-Schur-Concave case, the product of diagonal elements of MSE matrix is chosen as the performance metric, whose optimal solution is also the capacity maximizing one. From Fig.~\ref{fig:9} it can be concluded that with fixed modulation constellation, nonlinear transceivers enjoy better BER performance than linear transceiver designs even with A-Schur-Convex objective function. It can also be discovered that the transceiver design with THP performs better than that with DFE. This is because THP is performed at the transmitter and it can avoid error propagation effects compared with DFE. However DFE is performed at the receiver and the received signals are inevitably corrupted by noises.

When imperfect CSI is considered, robust designs are usually more preferable than their non-robust counterparts that simply take the estimated CSI as true CSI. With channel estimator errors, at each hop it holds that ${\bf{H}}_k={\bf{\hat H}}_k+\Delta{\bf{H}}_k$ and for simplicity it is assumed that for the three hop the elements of $\Delta{\bf{H}}_k$ are i.i.d. and have the same variance, i.e., 
\begin{figure}[!ht]
\centering
\includegraphics[width=.34\textwidth]{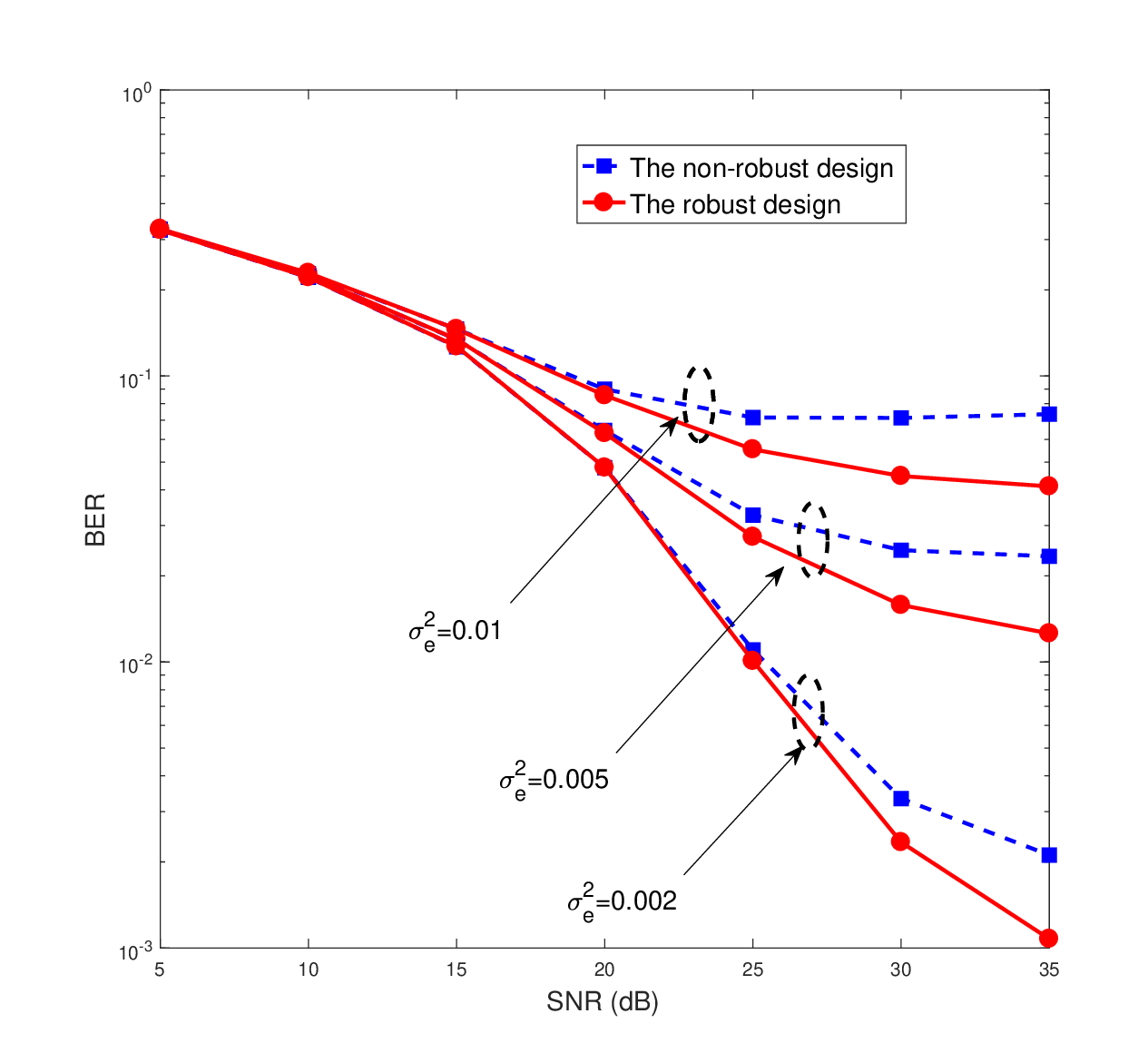}
\caption{The BERs of the nonlinear transceivers with THP under joint power constraints with $\tau_{k,\max}$=1.4 when CSI is imperfect.}\label{fig:10}
\end{figure} $\sigma_{e_1}^2=\sigma_{e_2}^2=\sigma_{e_3}^2=\sigma_{e}^2$.
When 16-QAM is adopted, in Fig.~\ref{fig:10} with different $\sigma_e^2$ the performance advantage of the robust transceiver with THP is demonstrated in terms of BER under joint power constraints with $\tau_{k,\max}=1.4$. For the curves in Fig.~\ref{fig:10}, Gray code is also used. It is worth noting that the robust design given by the remark at the end of Section VI has almost the same complexity as its non-robust counterpart.¡¡

\section{Conclusions}
\label{Conclusions}

In this paper, the transceiver designs for multi-hop AF MIMO relaying communications were investigated in depth under covariance shaping constraints. Under covariance shaping constraints a framework for transceiver designs was proposed in which various linear and nonlinear transceiver designs can be cast as a unified optimization problem. Afterward, it was discovered that the operations of AF MIMO relaying, THP and DFE can all be understood as a kind of matrix version weighting operations with different matrix version slopes and matrix version intercepts. For both the linear and nonlinear transceiver designs with various objective functions under pure shaping constraints or joint power constraints, the optimal solutions were derived in closed-form. Based on the derived solutions, it can be concluded that the proposed algorithm is well-suited for distributed implementation. Finally, the performance comparisons between different designs were given by the numerical results.

\appendices
\section{Fundamentals of Majorization Theory}
\label{Appendix_Majorization}

In this appendix, some fundamentals of majorization theory are introduced, which are the theoretical basis of our work. The interested readers are referred to the textbook \cite{Marshall2010} for more details.

\noindent \textbf{Definition 1.a:} For any ${\bf{z}} \in {\mathbb{R}}^{N}$, let ${{z}}_{[k]}$ denote the $k^{\rm{th}}$ largest element of ${\bf{z}}$, i.e.,
${{z}}_{[1]}\ge \cdots \ge {{z}}_{[N]}$. Given two vectors ${\bf{v}},{\bf{u}}$, the relationship that ${\bf{u}}$ majorizes ${\bf{v}}$ additively, i.e., ${\bf{v}}\prec_{+} {\bf{u}}$ is defined as
\begin{align}
\label{def_1} &\sum_{i=1}^k {{v}}_{[i]}\le \sum_{i=1}^k {{u}}_{[i]}, \  k=1,\cdots,N-1 \ \text{and}
\ \sum_{i=1}^N {{v}}_{[i]}= \sum_{i=1}^N {{u}}_{[i]}.
\end{align}

\noindent \textbf{Definition 1.b:} A function ${\phi}(\bullet)$ is  additively Schur-convex if and only if ${\bf{v}}\prec_{+} {\bf{u}}$ implies ${\phi}({\bf{v}})\le {\phi}({\bf{u}})$. Notice that ${\phi}(\bullet)$ is additively Schur-concave if and only if $-{\phi}(\bullet)$ is additively  Schur-convex.

\noindent \textbf{Definition 2.a:} For any ${\bf{z}} \in {\mathbb{R}}^{N}$, let ${{z}}_{[k]}$ denote the $k^{\rm{th}}$ largest element of ${\bf{z}}$, i.e.,
${{z}}_{[1]}\ge \cdots \ge {{z}}_{[N]}$. Given two vectors ${\bf{v}},{\bf{u}}$ with \textbf{nonnegative} elements, the relationship that ${\bf{u}}$ majorizes ${\bf{v}}$ multiplicatively, i.e., ${\bf{v}}\prec_{\times} {\bf{u}}$ is defined as
\begin{align}
\label{def_2} &\prod_{i=1}^k {{v}}_{[i]}\le \prod_{i=1}^k {{u}}_{[i]}, \  k=1,\cdots,N-1 \ \text{and}
\ \prod_{i=1}^N {{v}}_{[i]}= \prod_{i=1}^N {{u}}_{[i]}.
\end{align}

\noindent \textbf{Definition 2.b:} A function ${\phi}(\bullet)$ is  multiplicatively Schur-convex if and only if ${\bf{v}}\prec_{\times} {\bf{u}}$ implies ${\phi}({\bf{v}})\le {\phi}({\bf{u}})$. Notice that ${\phi}(\bullet)$ is multiplicatively Schur-concave if and only if $-{\phi}(\bullet)$ is multiplicatively  Schur-convex.

\noindent \textbf{Lemma 1:} If $\phi(\bullet)$ is a real-valued function defined on $\mathcal{D}= \{{\bf{z}}: z_1\ge \cdots \ge z_N\}$, ${\phi}(\bullet)$ is additively Schur-convex if and only if
  for all ${\bf{z}} \in \mathcal{D}$,
\[{\phi}(z_1,\cdots,z_{k-1},z_{k}-e, z_{k+1}+ e,z_{k+2},\cdots,z_N)\]
is decreasing in $e$ over the following defined regions
\begin{align}
& 0 \le e \ \text{and} \  z_{k}-e  \ge  z_{k+1}+ e   \ \ \text{for} \ \ k=1,\cdots,N-1.
\end{align}

\noindent \textbf{\textsl{Proof:}} The detailed proof can be found on Page 80 of the textbook \cite{Marshall2010}. $\blacksquare$

\noindent \textbf{Lemma 2:} If $\phi(\bullet)$ is a continuous real-valued function defined on $\mathcal{D}= \{{\bf{z}}: z_1\ge \cdots \ge z_N \ge 0\}$, ${\phi}(\bullet)$ is multiplicatively  Schur-convex if and only if
  for all ${\bf{z}} \in \mathcal{D}$,
\[{\phi}(z_1,\cdots,z_{k-1},z_{k}/e, z_{k+1}\times e,z_{k+2},\cdots,z_N)
\]is decreasing in $e$ over the following defined regions
\begin{align}
& 1\le e \ \text{and} \  z_{k}/e  \ge  z_{k+1}\times e   \ \ \text{for} \ \ k=1,\cdots,N-1.
\end{align}

\noindent \textbf{\textsl{Proof:}} The detailed proof can be found in Appendix~A in \cite{JSAC_Xing2012}. $\blacksquare$

\section{Derivation of ${\bf{U}}_{{\bf{F}}_k}$}
\label{Appendix_U}

At the beginning for convenience the column vectors of ${\bf{U}}_{{\bf{F}}_k}$ and ${\bf{U}}_{{\bf{R}}_{{\bf{s}}_k}}$ are  defined as ${\bf{u}}_{f_k,j}$'s and ${\bf{u}}_{s_k,j}$'s
i.e.,
\begin{align}
{\bf{U}}_{{\bf{F}}_k}&=[{\bf{u}}_{f_k,1} \  \cdots {\bf{u}}_{f_k,N_k} \cdots{\bf{u}}_{f_k,N_{T,k}}] \nonumber \\
{\bf{U}}_{{\bf{R}}_{{\bf{s}}_k}}&=[{\bf{u}}_{s_k,1} \  \cdots {\bf{u}}_{s_k,N_k} \cdots{\bf{u}}_{s_k,N}].
\end{align}Then in the following, the focus is how to prove ${\bf{u}}_{f_k,j}={\bf{u}}_{s_k,j}$ for $j=1,\cdots,N_k$. As ${\bf{u}}_{f_k,j}$ corresponds to the $j^{\rm{th}}$ largest eigenvalue of ${\bf{F}}_k{\bf{F}}_k^{\rm{H}}$, the following equality holds \cite[P.176]{Horn85}
\begin{align}
{\lambda}_1({\bf{F}}_k{\bf{F}}_k^{\rm{H}})=&\max_{{\bf{u}}^{\rm{H}}{\bf{u}}=1 } {\bf{u}}^{\rm{H}}{\bf{F}}_k{\bf{F}}_k^{\rm{H}}{\bf{u}} \nonumber \\
{\bf{u}}_{f_k,1}=&{\rm{argumax}}_{{\bf{u}}^{\rm{H}}{\bf{u}}=1 } \{{\bf{u}}^{\rm{H}}{\bf{F}}_k{\bf{F}}_k^{\rm{H}}{\bf{u}}\}
\end{align}based on which and together with the fact that ${\bf{F}}_k{\bf{F}}_k^{\rm{H}} \preceq {\bf{R}}_{{\bf{s}}_k}$ we will directly have the following inequality
\begin{align}
{\lambda}_1({\bf{F}}_k{\bf{F}}_k^{\rm{H}})= {\bf{u}}_{f_k,1}^{\rm{H}}{\bf{F}}_k{\bf{F}}_k^{\rm{H}}{\bf{u}}_{f_k,1}\le {\bf{u}}_{f_k,1}^{\rm{H}}{\bf{R}}_{{\bf{s}}_k}{\bf{u}}_{f_k,1}.
\end{align} Note that ${\lambda}_1({\bf{F}}_k{\bf{F}}_k^{\rm{H}})$ is also the largest the eigenvalue of ${\bf{R}}_{{\bf{s}}_k}$ and then the equality ${\lambda}_1({\bf{F}}_k{\bf{F}}_k^{\rm{H}})={\bf{u}}_{f_k,1}^{\rm{H}}{\bf{R}}_{{\bf{s}}_k}{\bf{u}}_{f_k,1}$ must hold. In other words, ${\bf{u}}_{f_k,1}$ is also the eigenvector corresponding the maximum eigenvalue for ${\bf{R}}_{{\bf{s}}_k}$, i.e.,
\begin{align}
{\bf{u}}_{f_k,1}={\bf{u}}_{s_k,1}.
\end{align}

Taking a further step, the second largest eigenvalue of ${\bf{F}}_k{\bf{F}}_k^{\rm{H}}$ satisfies \cite[P.177]{Horn85}
\begin{align}
{\lambda}_2({\bf{F}}_k{\bf{F}}_k^{\rm{H}})&= \max_{{\bf{u}}^{\rm{H}}{\bf{u}}=1,{\bf{u}} \perp {\bf{u}}_{f_k,1} } {\bf{u}}^{\rm{H}}{\bf{F}}_k{\bf{F}}_k^{\rm{H}}{\bf{u}} \nonumber \\
{\bf{u}}_{f_k,2}=&{\rm{argumax}}_{{\bf{u}}^{\rm{H}}{\bf{u}}=1,{\bf{u}} \perp {\bf{u}}_{f_k,1} } \{{\bf{u}}^{\rm{H}}{\bf{F}}_k{\bf{F}}_k^{\rm{H}}{\bf{u}}\}.
\end{align} Exploiting the facts that ${\bf{u}}_{f_k,2}$ is the second largest eigenvalue's eigenvector, i.e., ${\bf{u}}_{f_k,2}^{\rm{H}}{\bf{F}}_k{\bf{F}}_k^{\rm{H}}{\bf{u}}_{f_k,2}={\lambda}_2({\bf{F}}_k{\bf{F}}_k^{\rm{H}})$ and ${\bf{F}}_k{\bf{F}}_k^{\rm{H}} \preceq {\bf{R}}_{{\bf{s}}_k}$, we will directly have the following inequality
\begin{align}
{\lambda}_2({\bf{F}}_k{\bf{F}}_k^{\rm{H}})\le {\bf{u}}_{f_k,2}^{\rm{H}}{\bf{R}}_{{\bf{s}}_k}{\bf{u}}_{f_k,2}.
\end{align}It is worth noting that ${\lambda}_2({\bf{F}}_k{\bf{F}}_k^{\rm{H}})$ is also the second largest eigenvalue of ${\bf{R}}_{{\bf{s}}_k}$. Based on the fact that ${\bf{u}}_{f_k,2} \perp {\bf{u}}_{f_k,1}$, we have ${\bf{u}}_{f_k,2}^{\rm{H}}{\bf{R}}_{{\bf{s}}_k}{\bf{u}}_{f_k,2}\le {\lambda}_2({\bf{R}}_{{\bf{s}}_k})$. Similar to the previous logic for ${\bf{u}}_{f_k,1}$ the above equality must hold and therefore ${\bf{u}}_{f_k,2}$ is also the eigenvector corresponding the second largest eigenvalue for ${\bf{R}}_{{\bf{s}}_k}$, i.e.,
\begin{align}
{\bf{u}}_{f_k,2}={\bf{u}}_{s_k,2}.
\end{align}
Repeating this logic, it can be proved that ${\bf{u}}_{f_k,j}={\bf{u}}_{s_k,j}$ for
$j=1,\cdots,N_k$.

\section{Derivation of Optimal Structure of ${\bf{F}}_k$'s}
\label{Optimal_Structure}

In this section, the optimal solution of \textbf{Prob. 2} will be derived. By removing the final constraint, \textbf{Prob. 2} is relaxed to be the following much simpler one which is of the following form
\begin{align}
&{\textbf{Prob. 3:}}\nonumber \\
 &  \max_{\alpha,{\bf{F}}_k} \ \ \ \ \ \ \ \ \ \ \ \  \alpha \nonumber \\
& \ \ \ {\rm{s.t.}} \ \ \ \  {\bf{F}}^{\rm{H}}_k{\bf{H}}_{k}^{\rm{H}}
{\bf{R}}_{{\bf{n}}_k}^{-1}{\bf{H}}_k{\bf{F}}_k ={\alpha}{\bf{F}}^{\rm{H}}_{k,{\rm{PO}}}{\bf{H}}_{k}^{\rm{H}}
{\bf{R}}_{{\bf{n}}_k}^{-1}{\bf{H}}_k{\bf{F}}_{k,{\rm{PO}}}\nonumber \\
& \ \ \ \ \ \ \ \ \ \ \ {\rm{Tr}}({\bf{F}}_k{\bf{F}}_k^{\rm{H}})\le P_k.
\end{align}Based on the SVD
${\bf{R}}_{{\bf{n}}_k}^{-1/2}{\bf{H}}_k={\bf{U}}_{{\boldsymbol{\mathcal{H}}}_k}
{\Lambda}_{{\boldsymbol{\mathcal{H}}}_k}{\bf{V}}_{{\boldsymbol{\mathcal{H}}}_k}^{\rm{H}}$ with ${\Lambda}_{{\boldsymbol{\mathcal{H}}}_k} \searrow $, it has been shown in Appendix A in \cite{XingTSP2014} that the optimal solutions have the following structure
\begin{align}
\label{app_68}
{\bf{F}}_{k,{\rm{opt}}}={\bf{V}}_{{\boldsymbol {\mathcal H}}_k}{\boldsymbol \Lambda}_{{\bf{F}}_k}{\bf{U}}_{{\rm{Arb}},k}^{\rm{H}} \ \ {\rm with}  \ \  {\boldsymbol \Lambda}_{{\bf{F}}_k}^{\rm{T}}{\Lambda}_{{\boldsymbol{\mathcal{H}}}_k}^{\rm{T}} {\Lambda}_{{\boldsymbol{\mathcal{H}}}_k}{\boldsymbol \Lambda}_{{\bf{F}}_k} \searrow.
 \end{align}In the following we will show that the final constraint in \textbf{Prob. 2} will be automatically  satisfied.

In the following, firstly we will show that the optimal value of $\alpha$ satisfies $\alpha=1$. If $\alpha<1$, it is obvious that the computed ${\bf{F}}_k$ is not optimal as we can simply set ${\bf{F}}_k={\bf{F}}_{k,{\rm{PO}}}$ to achieve a better objective value without violating any constraint. Otherwise if $\alpha>1$ it contradicts with the fact ${\bf{F}}_{k,{\rm{PO}}}$ is Pareto optimal. In this case, as ${\bf{F}}^{\rm{H}}_k{\bf{H}}_{k}^{\rm{H}}
{\bf{R}}_{{\bf{n}}_k}^{-1}{\bf{H}}_k{\bf{F}}_k ={\alpha}{\bf{F}}^{\rm{H}}_{k,{\rm{PO}}}{\bf{H}}_{k}^{\rm{H}}
{\bf{R}}_{{\bf{n}}_k}^{-1}{\bf{H}}_k{\bf{F}}_{k,{\rm{PO}}}$ for the optimal values of $\alpha$ and ${\bf{F}}_k $, we can have
\begin{align}
\label{app_69}
&\lambda_i(\frac{1}{\alpha}{\bf{F}}^{\rm{H}}_k{\bf{H}}_{k}^{\rm{H}}
{\bf{R}}_{{\bf{n}}_k}^{-1}{\bf{H}}_k{\bf{F}}_k)\nonumber \\
&=\lambda_i
({\bf{F}}^{\rm{H}}_{k,{\rm{PO}}}{\bf{H}}_{k}^{\rm{H}}
{\bf{R}}_{{\bf{n}}_k}^{-1}{\bf{H}}_k{\bf{F}}_{k,{\rm{PO}}})\nonumber \\
&=\lambda_i(
{\bf{R}}_{{\bf{n}}_k}^{-1/2}{\bf{H}}_k{\bf{F}}_{k,{\rm{PO}}}
{\bf{F}}^{\rm{H}}_{k,{\rm{PO}}}{\bf{H}}_{k}^{\rm{H}}
{\bf{R}}_{{\bf{n}}_k}^{-1/2})\nonumber \\
&\le \tau_{k,\max} \lambda_i(
{\bf{R}}_{{\bf{n}}_k}^{-1/2}{\bf{H}}_k{\bf{H}}_{k}^{\rm{H}}
{\bf{R}}_{{\bf{n}}_k}^{-1/2})
\end{align}where the final inequality comes from the fact that ${\bf{F}}_{k,{\rm{PO}}}
{\bf{F}}^{\rm{H}}_{k,{\rm{PO}}} \preceq \tau_{k,{\rm{max}}}{\bf{I}}$. Substituting (\ref{app_68}) into (\ref{app_69}) we can prove that $\lambda_i(\frac{1}{\alpha}{\bf{F}}_k{\bf{F}}_k^{\rm{H}}) \le \tau_{k,\max}$. Note that at the optimum of \textbf{Prob. 3},  ${\rm{Tr}}({\bf{F}}_{k,,\rm{opt}}{\bf{F}}_{k,,\rm{opt}}^{\rm{H}})=P_k$ \cite{XingTSP2014} and if $\alpha>1$, we will have a new variable $\frac{1}{\sqrt{\alpha}}{\bf{F}}_{k,\rm{opt}}$ which satisfies all the constraints in \textbf{Prob. 2} but $\frac{1}{\alpha}{\rm{Tr}}({\bf{F}}_{k,\rm{opt}}{\bf{F}}_{k,\rm{opt}}^{\rm{H}})<P_k$. It means some power is not used and we can simply allocate it to the eigenmodels of $\frac{1}{\alpha}{\bf{F}}_{k,\rm{opt}}{\bf{F}}_{k,\rm{opt}}^{\rm{H}}$ whose power is smaller than the threshold. In other words, we can find a new ${\bf{F}}_k$ which satisfies all the constraints in \textbf{Prob. 2} makes ${\bf{F}}^{\rm{H}}_k{\bf{H}}_{k}^{\rm{H}}
{\bf{R}}_{{\bf{n}}_k}^{-1}{\bf{H}}_k{\bf{F}}_k \succeq {\bf{F}}^{\rm{H}}_{k,{\rm{PO}}}{\bf{H}}_{k}^{\rm{H}}
{\bf{R}}_{{\bf{n}}_k}^{-1}{\bf{H}}_k{\bf{F}}_{k,{\rm{PO}}}$. This conclusion contradicts with the fact that ${\bf{F}}_{k,{\rm{PO}}}$ is Pareto optimal. Finally, it can be concluded that ${\alpha}=1$. Together with (\ref{app_69}) it can be concluded that the relaxation of the final constraint is tight and \textbf{Prob. 2} and {\textbf{Prob. 3}} have the same optimal solutions. Therefore, (\ref{app_68}) is exactly the optimal structure of the optimal solutions to {\textbf{Prob. 2}}. Note that here we only focus on the case that for the Pareto optimal solution set the sum power constraint is always active. If this assumption is relaxed, a case may happen that the extra power cannot be allocated to the eigenmodels of $\frac{1}{\alpha}{\bf{F}}_{k,\rm{opt}}{\bf{F}}_{k,\rm{opt}}^{\rm{H}}$ because of the peak power constraints. However, in this case \textbf{Prob. 1} becomes much simpler as it is a special case with pure shaping constraints. We discover that the structures of the Pareto optimal solutions can still be written as (\ref{app_68}).

\section*{ACKNOWLEDGEMENT}
The authors would like to thank the anonymous reviewers for their valuable and
professional comments that have greatly improved the quality
of paper, especially for the work related to rank constraints.

\end{document}